 \documentclass[aps,prb,reprint,superscriptaddress]{revtex4-1}
\usepackage[dvipdfmx]{graphicx}
\usepackage{color}
\usepackage{gensymb}
\usepackage{ifthen}
\usepackage[version=3]{mhchem}
\usepackage{miller}
\usepackage{xspace}
\usepackage{textcomp}
\usepackage{nicefrac}
\usepackage{amsmath}

\usepackage{multirow}
\usepackage{longtable}

\usepackage{bm}

\usepackage[normalem]{ulem}

\newboolean{printSections} 
\setboolean{printSections}{true} 

\newcommand{\Cu}{\ce{Cu2OSeO3}\xspace}

\begin{document}


\title{Higher-order modulations in the skyrmion-lattice phase of \Cu}


\author{Johannes D. Reim}
\email{johannes.reim@rwth-aachen.de}
\author{Shinnosuke Matsuzaka}
\email{shinnosuke.matsuzaka.r2@dc.tohoku.ac.jp}
\author{Koya Makino}
\author{Seno Aji}
\author{Ryo Murasaki}
\author{Daiki Higashi}
\author{Daisuke Okuyama}
\affiliation{Institute of Multidisciplinary Research for Advanced Materials,
Tohoku University, 2-1-1 Katahira, Sendai 980-8577, Japan}

\author{Yusuke Nambu}
\affiliation{Institute for Materials Research, 
Tohoku University, 2-1-1 Katahira, Sendai 980-8577, Japan}
\affiliation{Organization for Advanced Studies, 
Tohoku University, 2-1-1 Katahira, Sendai 980-8577, Japan}
\affiliation{FOREST, Japan Science and Technology Agency,
Kawaguchi, Saitama 332-0012, Japan}

\author{Elliot P. Gilbert}
\author{Norman Booth}
\affiliation{Australian Centre for Neutron Scattering, Australian Nuclear Science and Technology Organization,
Kirrawee DC, New South Wales 2232, Australia}

\author{Shinichiro Seki}
\affiliation{RIKEN Center for Emergent Matter Science (CEMS), Wako, Saitama 351-0198, Japan}
\affiliation{Department of Applied Physics and Quantum Phase Electronics Center (QPEC), University of Tokyo, Tokyo 113-8656, Japan}

\author{Yoshinori Tokura}
\affiliation{RIKEN Center for Emergent Matter Science (CEMS), Wako, Saitama 351-0198, Japan}
\affiliation{Department of Applied Physics and Quantum Phase Electronics Center (QPEC), University of Tokyo, Tokyo 113-8656, Japan}

\author{Taku J Sato}
\affiliation{Institute of Multidisciplinary Research for Advanced Materials,
Tohoku University, 2-1-1 Katahira, Sendai 980-8577, Japan}


\date{\today}

\begin{abstract}
Using small angle neutron scattering, we have investigated higher-order peaks in the skyrmion-lattice phase of \Cu, in which two different skyrmion lattices, SkX1 and SkX2, are known to form. 
For each skyrmion-lattice phase, we observed two sets of symmetrically inequivalent peaks at the higher-order-reflection positions with the indices \hkl(110) and \hkl(200).
Under the condition where the SkX1 and SkX2 coexist, we confirmed the absence of the scattering at $\mathbf{Q}$ positions combining reflections from the two phases, indicating a significantly weak double-scattering component.
Detailed analysis of the peak profile, as well as the temperature and magnetic-field dependence of the peak intensity, also supports the intrinsic higher-order modulation rather than the parasitic double scattering.
The two higher-order modulations show contrasting magnetic-field dependence; the former \hkl(110) increases as the field is increased, whereas the latter \hkl(200) decreases.
This indicates that, in Cu$_2$OSeO$_3$, skyrmions are weakly distorted, and the distortion is field-dependent in a way that the dominant higher-order modulation switches from \hkl(110) to \hkl(200) under field.
Monte Carlo simulations under sweeping external magnetic field qualitatively reproduce the observed magnetic-field dependence, and suggests that the higher-order modulations correspond to the superlattices of weak swirlings appearing in the middle of the original triangular-latticed skyrmions.

\end{abstract}


\maketitle

\ifprintSections
\section{Introduction}
\fi
In recent years, research on chiral magnets has attracted considerable interest.
Among this class of materials, certain compounds exhibit topologically protected swirlings called skyrmions and are characterized by a topological quantum number~\cite{Bogdanov1989,Nagaosa2013,BackC2020}.
Besides their intriguing nature, these entities offer properties well suited for application in information technology~\cite{FertA2017}.
Triangular lattices formed from such skyrmions were first discovered in \ce{MnSi}~\cite{Muehlbauer2009} and subsequently in other compounds such as FeGe~\cite{Yu2011}, (Fe,Co)Si~\cite{Muenzer2010}, Cu$_2$OSeO$_3$~\cite{Seki2012,Adams2012}, Co$_8$Zn$_8$Mn$_4$~\cite{Tokunaga2015}, GaV$_4$S$_8$~\cite{KezsmarkiI2015}, and so on.
In bulk samples, this new structure was first observed exclusively in a small phase region at low temperatures and non-zero magnetic fields; however, several materials with more convenient stabilization conditions have been discovered recently, making its application in spintronics more likely\cite{Tokunaga2015,Karube2016}; this is especially so since techniques suitable for reading and writing have been proposed~\cite{Sampaio2013,Jiang2015,Fert2017,Hsu2017}, and external field control of skyrmions has become a likely possibility as well~\cite{JonietzF10,SchulzT12,JiangW17,LitziusK17,ZhangSL18,Okuyama19}.


Among various skyrmion-hosting compounds listed above, \Cu has attracted particular interest due to its multiferroic properties associated with its insulating nature~\cite{Seki2012_1,Seki2012_2}. 
The presence of a skyrmion lattice in \Cu has been shown using various techniques, including small angle neutron scattering (SANS)~\cite{Seki2012,Adams2012}. 
In \Cu, it has been known that two different skyrmion lattices (SkX1 and SkX2), which are rotated roughly 30\degree\ against each other, are stabilized at a certain ratio depending on a temperature ($T$) and magnetic-field ($H$) protocol used for the stabilization~\cite{Makino2017}.
Specifically, (1) cooling the sample within an applied field (FC) stabilizes the SkX2 skyrmion lattice, (2) SkX1 is predominant for field-warming (FW), and (3) following zero-field cooling (ZFC), the SkX1 and SkX2 skyrmion lattices are stabilized in coexistence and show no sign of relaxation behavior favoring one of these lattices at any point of the phase diagram. 

Using SANS, the long periodic modulation of the skyrmion lattice is observed as six-fold peaks in 2D intensity maps, originating from its triple-$\mathbf q$ nature.
Characteristic of the triple-$\mathbf q$ structure is the higher-order modulations arising from interference between the two or more fundamental modulations.
Observation of the higher-order modulation provides one form of direct evidence for the formation of a triple-$\mathbf q$ structure, as the counter possibility, namely the multi-domain single-$\mathbf q$ structure, does not give rise to the interference between higher-order modulations.
In addition, the phase relation of the three fundamental modulations may be known from the interference effect, which can be a direct confirmation of nontrivial topology~\cite{Adams2011}.

\begin{figure}
  \includegraphics[scale=0.5,trim=1.8cm 0cm 1.8cm 0cm]{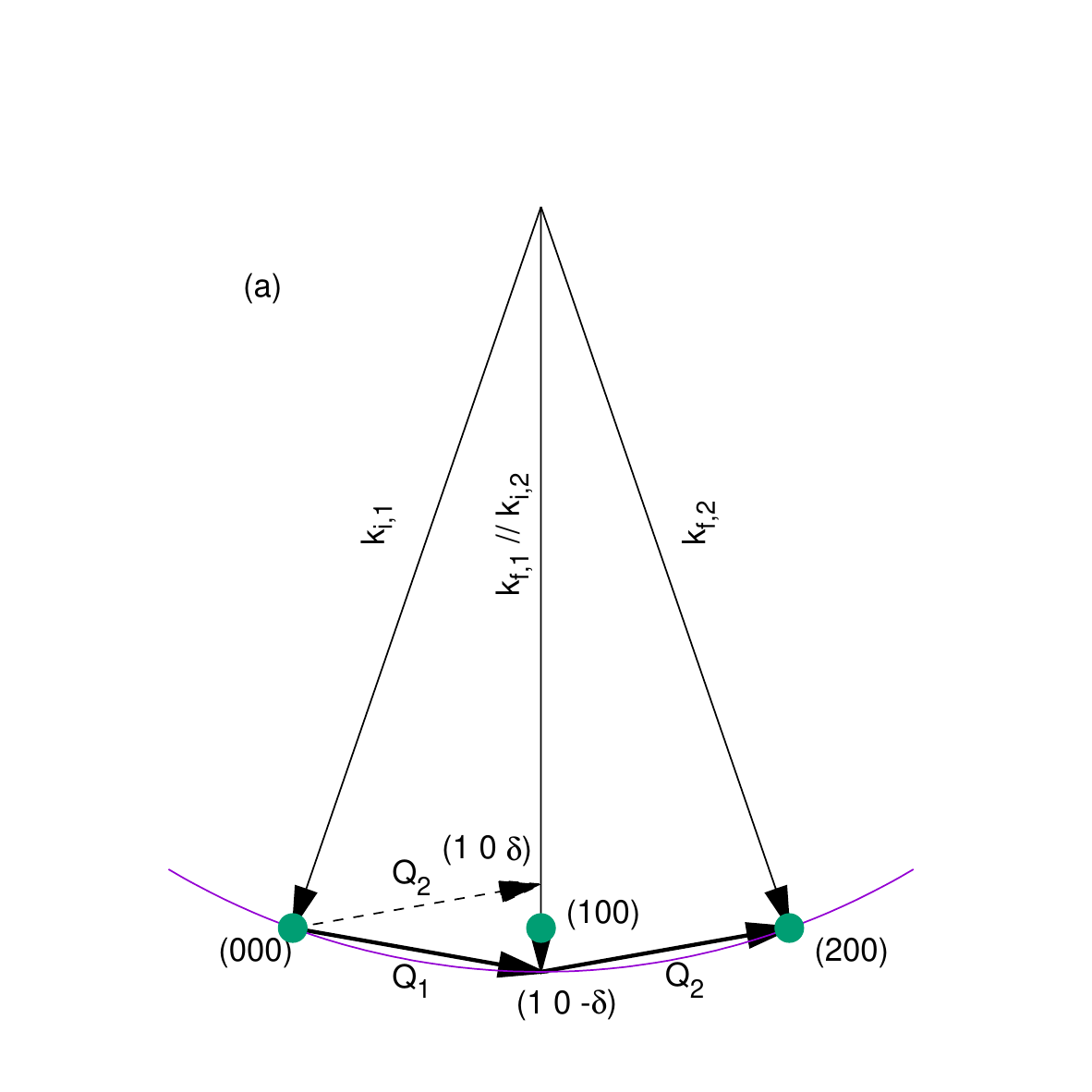}
  \includegraphics[scale=0.6,trim=2.5cm 1.5cm 2.5cm 2.5cm]{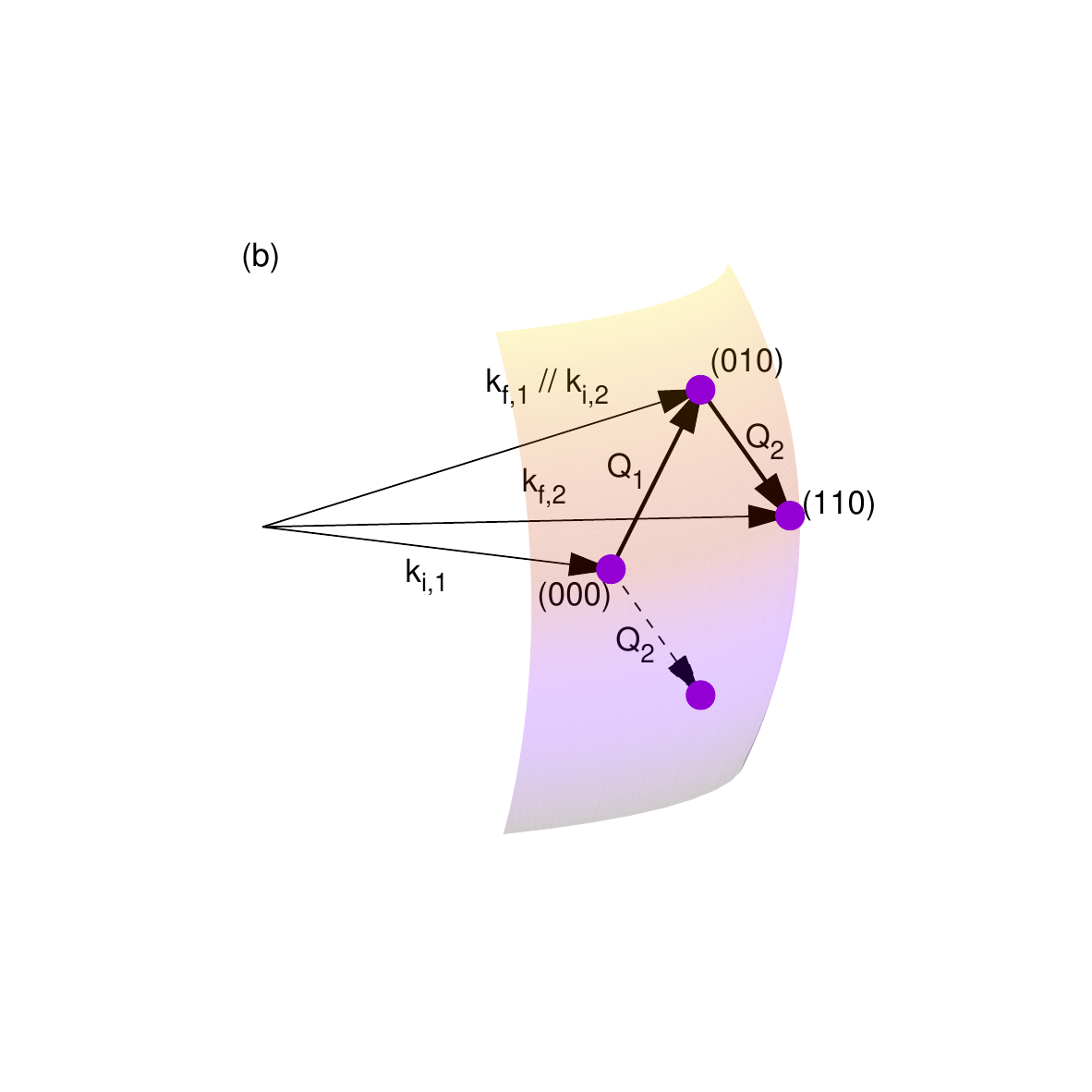}
  \caption{\label{fig:process} Schematic of the double scattering processes.
    The curvature of the Ewald sphere, as well as scattering angle $2\theta$, is significantly exaggerated to increase visibility.
    (a) 2D view from the vertical axis for the double scattering process for \hkl(200).
    Although the double scattering condition cannot be strictly satisfied in this case, due to the very small Ewald sphere curvature, the tail of the skyrmion Bragg peaks at $(1\ 0 \pm \delta)$ may result in finite double scattering probability at the \hkl(200) position.
    (b) 3D view for the double scattering process for \hkl(110).
    In this case, if both the \hkl(110) and \hkl(010) reflections [or \hkl(110) and \hkl(100) reflections if sample is tilted in the other direction] are simultaneously on the Ewald sphere, then a strict double scattering condition may be satisfied.
}
\end{figure}

Experimentally, however, there is a serious complication for the observation of the higher-order modulation in SANS, in which the scattering close to the origin is measured.
In this $Q$-range, the resulting Ewald-sphere is nearly flat, and more than one peak (or even all six peaks) may approximately fulfill the scattering condition at the same time, resulting in multiple scattering (scattering processes with a single neutron being scattered multiple times, $\mathbf{Q}_1, \mathbf{Q}_2, \ldots$), and, here, specifically its second order - double scattering~\cite{RenningerM1937,MoonRM1964,OkorokovAI2005,Adams2011}.
In general, the intensity and position of the double scattering may be similar to those of the higher-order scattering.
Two particularly important cases of the double scattering processes in the SANS setup are schematically shown in  Fig.~\ref{fig:process}.
In the first case, a neutron is scattered by the same first-order peak twice [cf. Fig.~\ref{fig:process}(a,b)], whereas the neutron can be scattered from another first-order peak subsequently in the second case [cf. Fig.~\ref{fig:process}(c)].
Those two scattering paths can result in misleading scattering intensity at higher-order positions. 


To date, there have been a limited number of experiments performed aiming at detecting the higher-order modulation in the skyrmion-lattice phase; a pioneering example may be the one on MnSi~\cite{Adams2011}.
In the experiment, the higher-order modulation was distinguished from the double scattering by performing Renninger scans~\cite{RenningerM1937}, where the sample is rotated in such a way that the scattering condition for a higher-order peak position remains fulfilled, and the one for the first order peaks is not.
However, such scans can generally be challenging to be implemented on SANS instruments as the sample has to be rotated around the axis defined by the $\mathbf Q$ position of the respective higher-order peak. 
Furthermore, at each rotation step, the statistics have to be sufficient to detect the weak higher-order scattering.

In this work, we have utilized a different approach to distinguish the intrinsic higher-order modulations from double scattering artifacts.
Specifically, we scrutinized extinction rules, peak profiles for sample $\theta$-rotation, and magnetic-field and temperature dependence of the scattering intensity at the first-order and higher-order positions.
All the results consistently indicate that the dominant contribution to the reflection intensity at the higher-order positions are intrinsic to \Cu.
Further support was obtained from numerical Monte Carlo simulation of a simple two-dimensional (2D) square lattice with competing ferromagnetic and Dzyaloshinskii-Moriya (DM) interactions under sweeping magnetic field; the magnetic-field dependence of the higher-order intensity is qualitatively reproduced.
%
%

\ifprintSections
\section{Experimental}
\fi
To investigate the structure of the skyrmion lattice in \Cu, small angle neutron scattering experiments were performed on a high quality single crystal.
The crystal with approximate dimensions of 8\,mm $\times$ 5\,mm $\times$ 3\,mm  has been synthesized using the chemical vapor transport method \cite{Miller2010}. The neutron experiments were conducted at the beam line QUOKKA located at the OPAL reactor of the Australian Nuclear Science and Technology Organisation, Australia~\cite{WoodK18}. 
To achieve the necessary resolution and sufficient neutron flux, a wavelength $\lambda \sim 5$\,\AA\ was selected using a neutron velocity selector, with wavelength distribution d$\lambda/\lambda \sim 10$\% in combination with a 2D detector. 
The skyrmion lattice was stabilized using a closed-cycle refrigerator and 
an external 5\,T horizontal field superconducting magnet with its field aligned parallel to the incident neutron beam $\mathbf{k}_i$. The sample was oriented with its crystallographic \hkl[110] axis approximately parallel to $\mathbf{k}_i$ (slightly misaligned by $\sim 6$\textdegree\ in the horizontal scattering plane and $\sim5$\textdegree\ against the plane) and crystallographic \hkl[001] along the horizontal direction.
The sample rotation axis for the angle $\theta$ is normal of the horizontal scattering plane (rocking scan). 
The scattering background was estimated using the SANS pattern measured at the paramagnetic temperature 60\,K, 
and subtracted from data shown here.
The wavelength used in the present experiment allows to measure the scattering intensity of all the six-fold symmetry-equivalent reflections from the skyrmion lattice, appearing at $|\mathbf{Q}| \simeq 0.01$~\AA$^{-1}$, simultaneously, yet at different intensities.
Throughout this manuscript, we use indices in the parentheses \hkl(hk0) to refer to the skyrmion lattice reflections; the \hkl(100) reflection corresponds to the first order peak appearing at $|\mathbf{Q}| \simeq 0.01$~\AA$^{-1}$ approximately along the horizontal $Q_x$ axis in two-dimensional SANS patterns.


\ifprintSections
\section{Data analysis}
\fi
\label{sec:method}

\subsection{Fitting model and procedure}
\label{subsec:model}
In the investigation of the first order peaks of the skyrmion lattice \hkl(100), data analysis was often simplified by integrating the data over a certain $Q$-range.
This reduces the 2D data to a 1D azimuthal variation, which still contains all relevant information.
While in principle this could be done in the present analysis on the potential higher-order and double scattering, practically the simplification encounters a problem.
In typical intensity maps obtained in the present study, besides the first-order peak indexed as \hkl(100) and its symmetry equivalents, additional peaks are visible at higher Q, if viewed on a logarithmic scale [see Fig. \ref{fig:map}(a)]. 
These can be indexed as symmetry equivalents of \hkl(110) and \hkl(200) respectively.
Because of the finite peak widths, a separation of them is difficult, and thus the simple integration method used for the analysis of the first-order peaks is not applicable.
This becomes a more serious issue for the skyrmion-lattice formed after the FW and ZFC runs; due to the coexistence of the two different skyrmion-lattice phases (SkX1 and SkX2), the higher-order reflections appear much closer [see Figs. \ref{fig:mapZFC}].
Furthermore, the intensity for the higher-order reflections is significantly weaker, which requires rigorous analysis to reliably extract the reflection intensity from the experimental intensity maps.
Consequently, the scattering patterns are fitted in 2D to extract intensities, positions and peak widths; the peak shape in 2D allows for reliable fitting even when the peaks overlap.

Each peak will be approximated by a 2D Gaussian function defined as:
\begin{equation}
G(\mathbf{Q},I_0, \mathbf{Q}_0,\sigma_{\mathbf Q},\sigma_\Phi) 
= I_0 e^{-\left( \frac{(Q_x-Q_{0x})^2}{2\sigma_{\mathbf Q \parallel}^2} + \frac{(Q_y-Q_{0y})^2}{2\sigma_{\mathbf Q\perp}^2}\right)},
\end{equation}
with the amplitude $I_0$, the center position $\mathbf{Q} = (Q_{x0},Q_{y0})$, the peak widths $\sigma_{\mathbf{Q}\parallel}$ along the radial ($\mathbf Q$) and $\sigma_{\mathbf{Q}\perp}$ perpendicular to $\mathbf{Q}$ ({\it i.e.} along the azimuthal direction.) 
Due to the curvature of the Ewald sphere, the intensities of peaks at opposite $\mathbf Q$ are not necessarily identical, which is why the intensity of each peak is fitted independently.
Regarding the position, for each peak set (SkX1 or SkX2) independently a 6-fold symmetry is assumed, with the higher-order-peak positions being calculated from the first-order ones. 

Aside from the peaks, a significant diffuse scattering is present, which is strongest at $Q_{(100)}$, invariant under the azimuthal angle \cite{Makino2017} and falls off exponentially to higher $Q$.
This will be modeled using a 1D Gaussian modified with an exponential function:
\begin{equation}
        \mathrm{BG}(Q, I_0, \mu, \sigma, \lambda) = 
        I_0 \frac{\lambda}{2} e^{\frac{\lambda}{2}\cdot(2\mu+\lambda\sigma^2-2Q)}
        \mathrm{erfc}(\frac{\mu+\lambda\sigma^2-Q}{\sqrt{2}\sigma}),
\end{equation}
with the amplitude $I_0$, mean $\mu$, standard deviation $\sigma$, exponential rate $\lambda$ 
and the complementary error function:
\begin{equation}
 \mathrm{erfc}(x) = \frac{2}{\sqrt{\pi}} \int_x^\infty e^{-t^2} \mathrm{d}t.
\end{equation}
The background at high $Q$ is weak and fluctuating. 
Thus, only the peaks are fitted in 2D; from the model and the data, a pseudo powder pattern is subsequently calculated and the background is fitted to the difference in 1D. 
This process is repeated until convergence is achieved.  
Assuming SkX2 is stabilized exclusively, e.g. for the FC protocol, the equation for the full model is:
\begin{eqnarray}\label{eq:fit_2D}
I_\mathrm{2D}(\mathbf{Q}) &=& \sum_{n=1}^6 G(\mathbf{Q},I_{(100),n}, \mathbf{Q}_{0,n},\sigma_{\mathbf Q\parallel},\sigma_{\mathbf Q \perp})  \nonumber\\
&&+\begin{aligned}
        \sum_{n=1}^6 G(\mathbf{Q},I_{(110),n}, \mathbf{Q}_{0,n}+\mathbf{Q}_{0,n+1}, \sigma_{\mathbf Q \parallel},\sigma_{\mathbf Q \perp})
    \end{aligned}
   \nonumber \\
&&+\sum_{i=n}^6 G(\mathbf{Q},I_{(200),n}, 2\mathbf{Q}_{0,n},\sigma_{\mathbf Q \parallel},\sigma_{\mathbf Q \perp})  \nonumber \\
&&+ \mathrm{BG}(Q, I_0, \mu, \sigma, \lambda)
\end{eqnarray}
with $\mathbf{Q}_{0,n} = q[\cos(n\pi/3 +\Phi_0), \sin(n\pi/3 +\Phi_0), 0]$ where $\Phi_0$ describes the rotation of the lattice and $q$ is the length of the magnetic modulation vector.
For SkX1 or both lattices superimposed, the corresponding 2D Gaussians are added. Aside from the position in Q, the fit parameter for SkX1 and SkX2 are completely independent.
In general, this procedure results in a reasonable comparison even for the weak peaks - \hkl(110) and \hkl(200) -  and when SkX1 and SkX2 are superimposed, in 2D and in the pseudo powder (see sFigs.~2-11 in supplemental materials).
For consistency, the same procedure is used for all data sets. 
When only the first order peaks are of interest, the higher-order peaks are omitted from the model.

\subsection{First-order peak structure along the out-of-scattering-plane direction}
In contrast to the sharp feature of the first order Bragg peaks in the 2D scattering plane, these peaks exhibit elongated and structured peak profiles in the rocking scans along the out-of-scattering-plane ($\theta$) directions; indeed, the first-order peaks (and also higher-order peaks) show two broad peaks as a function of $\theta$, as will be described later.
To model this peak profile along the $\theta$ direction, we used the following resolution-convoluted two 1D Gaussians with a constant background:
\label{subsec:structure}
\begin{equation}\label{eq:fit_mosaicity}
 \begin{aligned}
 I_{n}(\theta) =& 
R(\theta)*\sum_{i=1}^2\left[G(\theta;\theta_{0,i}+\delta_n,I_{0,i},\sigma_i)\right] \\& + \mathrm{const.} 
 \end{aligned}
\end{equation}
The fitting was performed only for the two horizontal peaks p2 and p5 in Fig.~\ref{fig:map}(c) or $n \in \{2,5\}$.
The position $\theta_{0,i}$, amplitude $I_{0,i}$ and standard deviation $\sigma_i$ are the same for both peaks.
The resolution function $R(\theta)$ is obtained from the out-of-plane scans for the helical peaks.
Here, the 1D Gaussian function $G(\theta;\theta_0,I_0,\sigma) = I_0/(\sqrt{2\pi}\sigma)\mathrm{exp}\left[-(\theta-\theta_0)^2/2\sigma^2\right]$ is used and $\delta_n$ is the $\theta$-shift for the $n$-th peak, measured from the peak $\theta$ position for the $n = 2$ peak, i.e. $\delta_2 = 0$.

\subsection{Calculating double scattering}\label{subsection:doublescattering}
Double scattering stems from two elastic scattering processes (first: $\mathbf{Q}_1 = \mathbf{k}_{{\rm f},1} - \mathbf{k}_{{\rm i},1}$ and second: $\mathbf{Q}_2 = \mathbf{k}_{{\rm f},2} - \mathbf{k}_{{\rm i},2}$ with incident $\mathbf{k}_{{\rm i},m}$ and scattered $\mathbf{k}_{{\rm f},m}$ wavevectors), with their scattering vectors pointing to the same or neighboring first order peaks (relevant double scattering).
While the probability for the first scattering process with $\mathbf{Q}_1$ performed at sample rotation $\theta_1$ is proportional to the measured first-order reflection intensity, the probability for the second process should be different as the incident $\mathbf{k}_{{\rm i},2}$ of the second process is not parallel to $\mathbf{k}_{{\rm i},1}$ of the first process; $\mathbf{k}_{{\rm i},2} = \mathbf{k}_{{\rm i},1}+\mathbf{Q}_1$.
Nonetheless, the reflection $\mathbf{Q}_2$ for the second process is readily known from geometrical constraints for each double scattering process.
Hence, by assuming that all the first-order reflections have the same peak profiles along the $\theta$ direction, by taking account of the change of incident neutron direction between the first and second process, we may estimate the scattering probability for the second process, and accordingly, the total scattering probability, from the first order probability.

In practice, the measured profile function  $I_n(\theta)$ for the first order peak is converted $I(Q_z)$, a function of $Q_z$, and used as a scattering probability for the first-order process.
The second process probability is assumed to be $I_{n'}(Q_z')$, where $Q_z'$ is the position where the new Ewald sphere intersects the $\theta$ scan locus of the corresponding reciprocal-lattice vector (i.e. $n'$ corresponds to the number index of the first-order reflection appearing at $\mathbf{Q}_2$.)
The double scattering probability may be obtained by the product of $I_n(Q_z)I_{n'}(Q_z')$, where $n$ and $n'$ are the reflections which fullfill the double scattering condition for a specific higher-order position.

\ifprintSections
\section{Experimental results}
\fi

\subsection{Appearance of the higher-order reflections}
\begin{figure*}
\includegraphics{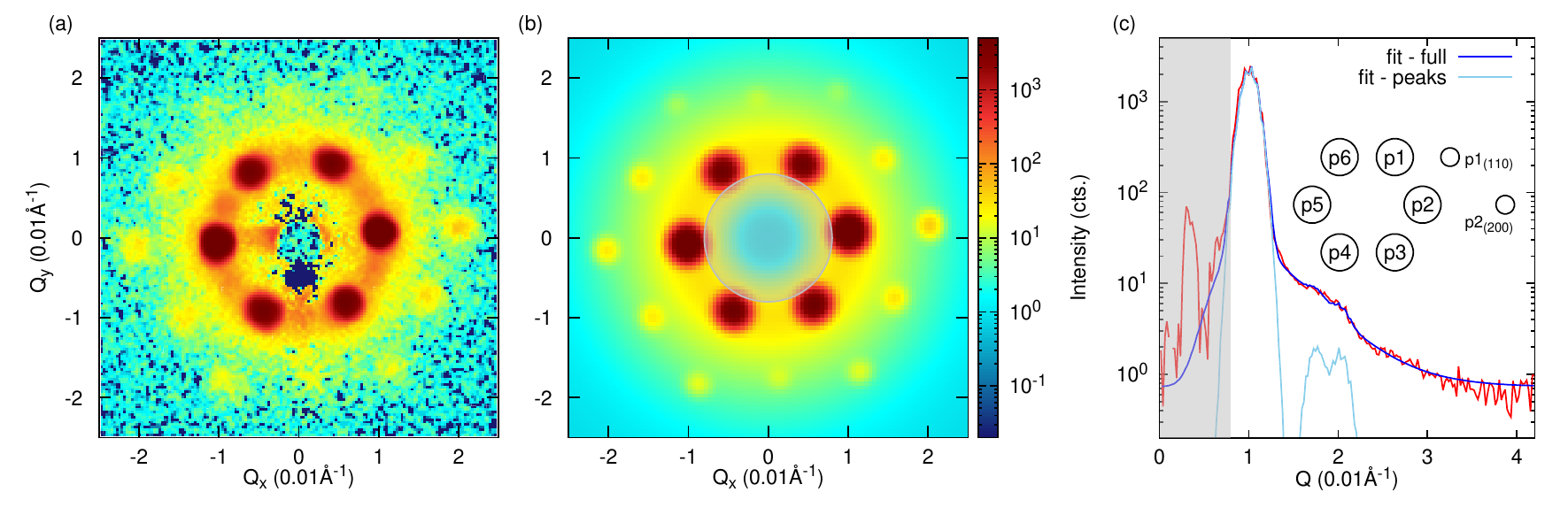}%
\caption{\label{fig:map} (a) 2D intensity map measured in the skyrmion-lattice phase (SkX2) stabilized at $T = 56.75$~K and $\mu_0 H = 17.4$~mT using the FC protocol.
  The intensity is shown on a logarithmic scale.
  (b) Reconstruction of the fit results based on the model defined by Eq. \ref{eq:fit_2D}.
  (c) The pseudo powder spectrum calculated from the 2D scattering data (red line), as well as that obtained from the 2D fit result with background function (dark blue line) and without the background (light blue line).
  The low-$Q$ gray region, where contamination due to the direct beam was found, was excluded in the fitting.
  The inset shows the definition of the first-order and higher-order peak number used in the main text.
}
\end{figure*}

\begin{figure*}
\includegraphics{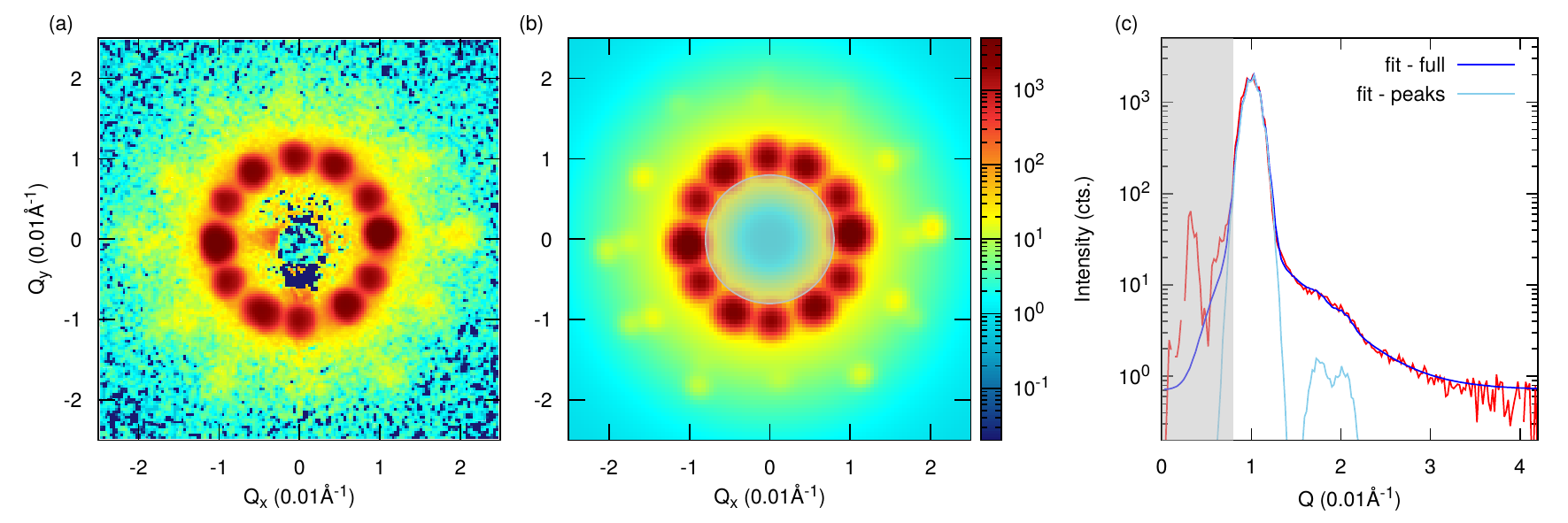}%
\caption{\label{fig:mapZFC}
  (a) 2D intensity map measured in the coexisting SkX1 and SkX2 phases stabilized at $T = 56.75$~K and $\mu_0 H = 17.4$~mT using the ZFC protocol.
  The intensity is shown on a logarithmic scale.    
  (b) Reconstruction of the fit results based on the model consisting of two sets of six-fold peaks, each of which is defined by Eq. \ref{eq:fit_2D}.
  (c) The pseudo powder spectrum calculated from the 2D scattering data (red line), as well as that obtained from the 2D fit result with background function (dark blue line) and without the background (light blue line).
  The low-$Q$ gray region, where contamination due to the direct beam was found, was excluded in the fitting.
}
\end{figure*}

To investigate supposedly weak intensity at the higher-order-reflection positions, we obtained SANS patterns at representative $H$ and $T$ points with very long exposure time, typically one hour or more.
The data shown in this study were originally obtained for the time-relaxation measurement reported in Ref.~[\onlinecite{Makino2017}].
Except for the first few minutes where a strong relaxation effect was observed, the SANS patterns are time-independent, and thus are integrated to obtain sufficiently high statistics.
Figure~\ref{fig:map}(a) shows the 2D SANS scattering intensity map on a logarithmic scale measured at $T = 56.75$~K and $\mu_0 H = 17.4$~mT using the FC protocol.
Clear formation of the SkX2 phase is observed from the appearance of the first-order six-fold reflections.
In addition, we can recognize finite scattering intensity at both the \hkl(110)- and \hkl(200)-type higher-order positions.
The 2D Gaussian fitting was performed for this scattering pattern, and the fitting result is shown in Fig.~\ref{fig:map}(b).
The characteristic features of the SANS pattern, such as peak intensity, position and shape, are well reproduced by the Gaussian fitting.
Shown in Fig.~\ref{fig:map}(c) is the azimuthally integrated (pseudo powder) diffraction patterns obtained from the experimental data (solid red line), and corresponding one obtained from the 2D Gaussian fitting result.
It can be seen that 2D Gaussian fitting quantitatively reproduces the experimental observation, indicating the reliability of the parameters obtained in the fitting.

It is known that coexistence of two different types of skyrmion-lattice phases is realized under the ZFC condition~\cite{Makino2017}.
Shown in Fig.~\ref{fig:mapZFC} is the SANS pattern measured at $T = 56.75$~K and $\mu_0 H = 17.4$~mT with the ZFC protocol.
Clearly, there appear twelve first-order reflections, indicating that the other SkX1 phase coexists with the SkX2 phase observed in the FC condition.
The 2D Gaussian fitting was similarly performed, and a successful result was obtained as shown in Fig.~\ref{fig:mapZFC}(b).
It can be clearly seen that the higher-order peaks of the indices \hkl(110) and \hkl(200) for the SkX2 phase, as well as those with index \hkl(110) for the SkX1 phase, are found in the ZFC SANS pattern.
It should be noted that there is no scattering intensity at the position \hkl(100)$_{\rm SkX1} +$\hkl(100)$_{\rm SkX2}$, i.e. a combination of magnetic modulations from the two different phases.
Since the double scattering should be equally possible for this case, the absence of the scattering intensity at \hkl(100)$_{\rm SkX1} +$\hkl(100)$_{\rm SkX2}$ unambiguously indicates that the parasitic double scattering is quite weak, compared to the intrinsic higher-order contribution in Cu$_2$OSeO$_3$.

\subsection{Temperature and magnetic field dependence of the higher-order intensity}

\begin{figure}
\includegraphics{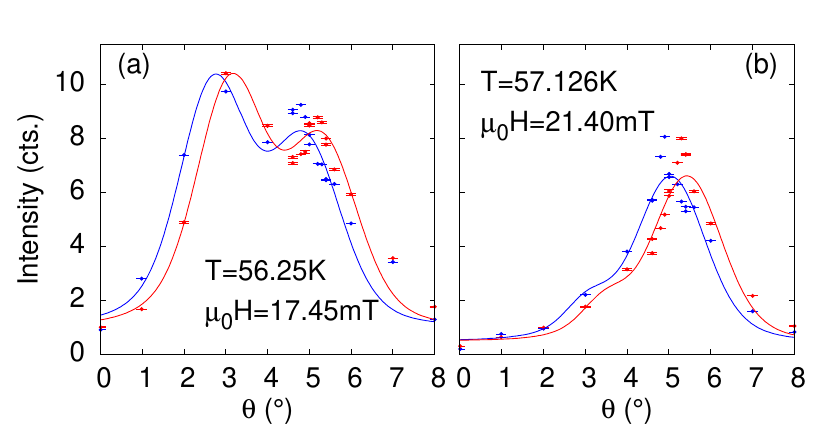}%
\caption{\label{fig:mosaicity_fitting}
  The sample $\theta$-rotation scans for the two first-order peaks, p2 (blue) and p5 (red).
  Scans were performed (a) at $T = 56.25$~K and $\mu_0 H = 17.45$~mT, and (b) at $T = 57.126$~K and $\mu_0 H = 21.4$~mT.
  Fit results to Eq.~\ref{eq:fit_mosaicity}) are shown by the solid lines.
  Clearly, two peaks were observed in the out-of-scattering-plane ($\theta$) direction.
}
\end{figure}
\begin{figure}[h]
\includegraphics{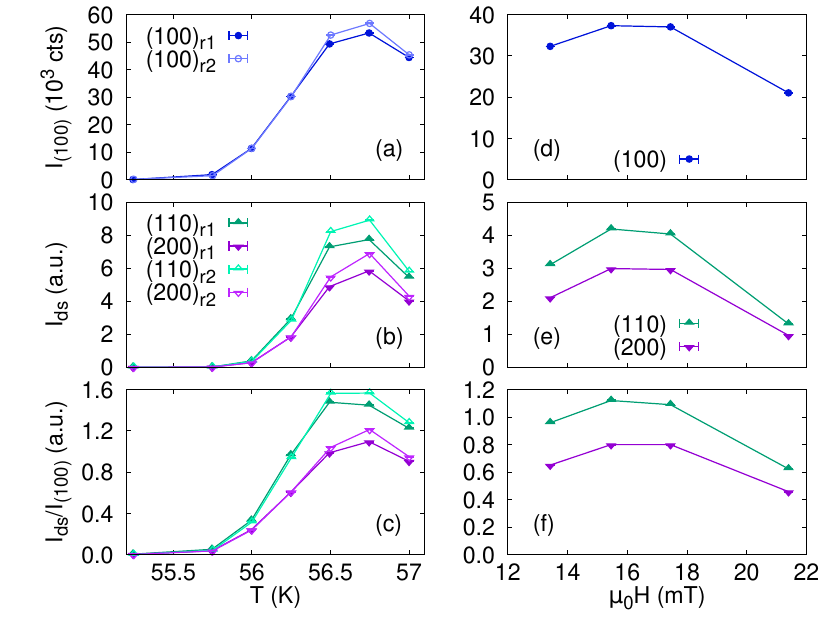}%
\caption{\label{fig:double_scattering} Results for the calculation of the double scattering probabilities based on the first order peak intensities measured in rocking scans.
For all rocking scans the FC protocol was used; thus all intensities stem from the SkX2 skyrmion lattice.
The temperature dependence (two runs, r1 and r2, have been performed) was measured at $\mu_0 H=17.45$\,mT (a-c) and the magnetic field dependence at $T=57.125$\,K (d-f). 
Here, the intensity of the six corresponding peaks is summed.
The calculated double scattering (b,e) follows directly the intensity of the first order peaks (a,d).
The same applies to the relation between double scattering and first order intensity (c,f).}
\end{figure}

\begin{figure}[h]
\includegraphics{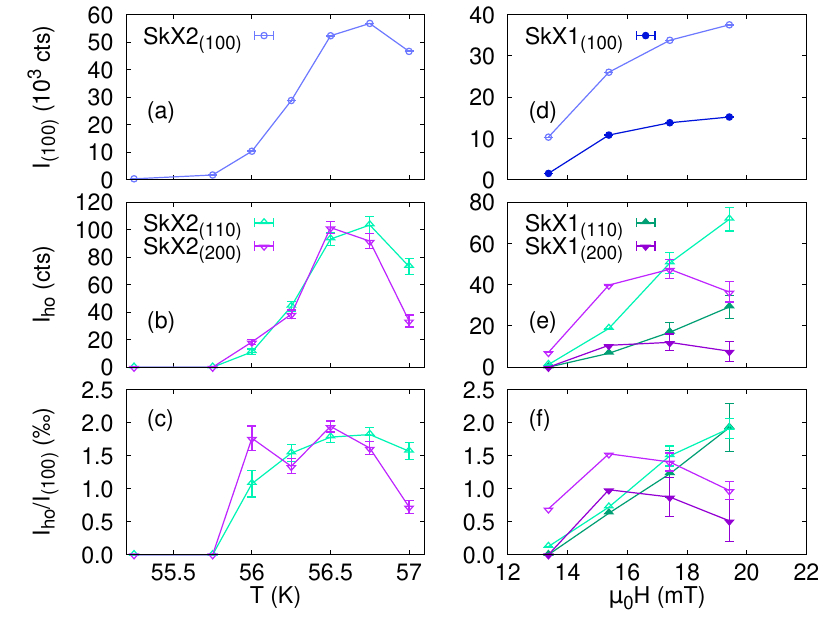}%
\caption{\label{fig:integ_intensity} Comparing the summed intensity of the first order peaks $I_\textrm{(100)}$ and the peaks at higher-order positions $I_\textrm{ho}$.
For the temperature dependence at $\mu_0 H=17.45$\,mT the FC protocol was used (a-c) and the ZFC protocol for the magnetic field dependence at $T=56.75$\,K (d-f). 
In the latter case in addition to SkX2 also the SkX1 skyrmion lattice was stabilized.
A strong dependence is visible for the higher-order peaks on both temperature (b) and magnetic field (e) apparently different for \hkl(110) and \hkl(200). 
With increasing temperature the \hkl(200) peaks decrease more quickly close to the phase transition temperature compared to \hkl(110). 
Normalizing by the first order intensity, $I_\textrm{ho}/I_\textrm{(100)}$ shows a deviation from the dependence of the first order peaks.}
\end{figure}

Intrinsically, the first-order and higher-order components of the skyrmion-lattice should show different dependence on the temperature and magnetic field~\cite{Adams2011}.
On the other hand, if the first-order intensity well localizes in the $\mathbf Q$-space, then the double scattering intensity should be proportional to the square of the first-order intensity.
Therefore, by comparing the $H$ and $T$ dependence of the first-order and higher-order peaks, one may distinguish the origin of the scattering at the higher-order positions.
However, in reality, there is a possibility that peak profiles for the out-of-plane ($\theta$) direction depend on $T$ and $H$.
This may result in different $T$ and $H$ dependence for the first-order and double-scattering intensities.
Below, we first confirm that, in the present experimental situation, the double scattering probability closely resembles that of the first-order scattering, even after taking care of the $T$ and $H$ dependence of the peak profile.
Then, we will show experimentally that the scattering intensities at the \hkl(100), \hkl(110) and \hkl(200) positions all show distinct $T$ and $H$ dependence.

The out-of-plane profile for the first-order reflections was measured by performing the sample $\theta$-rotation scans with a short exposure (15\,s); this provides sufficient statistics for the 2D fitting procedure when only the first order peaks are considered.
The scans were performed within the skyrmion phase at several temperatures under $\mu_0H=17.5$\,mT, and four different fields at $T=57.125$\,K, using the FC protocol stabilizing SkX2 exclusively. 
Each scattering map was well fitted by the 2D Gaussian model, and intensity obtained for the p2 and p5 peaks is plotted in terms of the angle $\theta$ in Fig.~\ref{fig:mosaicity_fitting}(a and b).
The two broad peaks are clearly seen in the figure, indicating the existence of two skyrmion-lattices, which are slightly tilted with respect to each other: SkX2$^\prime$ (at low $\theta$, $i=1$) and SkX2$^{\prime\prime}$ (at high $\theta$, $i=2$).
We find that the two broad peaks exhibit an intriguing temperature and magnetic field dependence, which will be explained in Appendix~\ref{appendix:rotationscans}.

From the obtained out-of-plane peak profile, we estimate the temperature and field dependence of the double-scattering intensity (probability) using the method described in subsection \ref{subsection:doublescattering}.
In Fig.~\ref{fig:double_scattering} the temperature (left column) and magnetic field (right column) dependence of the calculated double scattering intensity $I_\mathrm{ds}$ is displayed. 
It can be seen that the double scattering expected at the (110) and (200) positions [Fig. \ref{fig:double_scattering} (b and e)] closely follows the intensity of the first order peaks $I_{(100)}$. 
Even the ratio $I_\mathrm{ds}/I_{(100)}$ [Fig. \ref{fig:double_scattering} (c and f)] shows a very similar dependence indicating that both scattering processes in the configuration here are nearly identical. This is not unexpected, as the propagation vector $q$ is very small compared to the incident wave vector $k_{i,1}$.

To compare, the 2D SANS patterns were measured with a longer exposure (1800~s) at a fixed sample $\theta$-rotation angle.
The temperatures and fields were selected so that they correspond to those used for the $\theta$-scans of the first-order peak. 
For the temperature dependence at $\mu_0 H=17.45$\,mT, the skyrmion lattice was stabilized using the FC protocol, while for the magnetic field dependence at $T=56.75$\,K the ZFC protocol was used.
The 2D fitting procedure (described in Subsection~\ref{subsec:model}) was applied for the obtained intensity maps, and the $T$ and $H$ dependence of the first order intensity $I_{100}$, as well as those at the higher-order positions $I_{110}$ and $I_{200}$ was obtained.
Shown in Fig.~\ref{fig:integ_intensity}(a and d) is the resulting temperature and field dependence of the first order peaks for the SkX2 and SkX1 phases, respectively, whereas those shown in Fig.~\ref{fig:integ_intensity}(b and e) are the results for the higher-order reflections.
For the magnetic field dependence shown in Fig.~\ref{fig:integ_intensity}(e), while the \hkl(110) intensity from both the SkX1 and SkX2 phases increases monotonically with increasing field, the \hkl(200) intensity shows an initial increase, and then weakly decreasing behavior.
This nontrivial field dependence becomes clear when the ratio $I_{110}/I_{100}$ or $I_{200}/I_{100}$ is plotted as in Fig.~\ref{fig:integ_intensity}(f); all \hkl(100), \hkl(110), and \hkl(200) show different field dependences.
The scattering intensity for the first and higher-order peaks shows relatively similar temperature dependence [Fig.~\ref{fig:integ_intensity}(b)], however the ratio $I_{110}/I_{100}$ and $I_{200}/I_{100}$ shows weaker and flatter temperature dependence, indicating that the difference may also exist for the temperature dependence.
Such distinct $H$ and $T$ dependence for the first- and higher-order scattering intensity was found for all three protocols [see Fig.~\ref{fig:integ_intensity}(f), as well as sFig.~1(c,f) in supplemental materials].
As noted earlier in this subsection, the temperature and field dependence for the double scattering intensity should be closely related to those of the first-order reflections (see Fig. \ref{fig:double_scattering}).
Therefore, the significantly distinct field dependence between the observed first-order and higher-order intensity also supports the intrinsic origin of the higher-order contribution.

\subsection{Out-of-plane profile}
\begin{figure}
\includegraphics{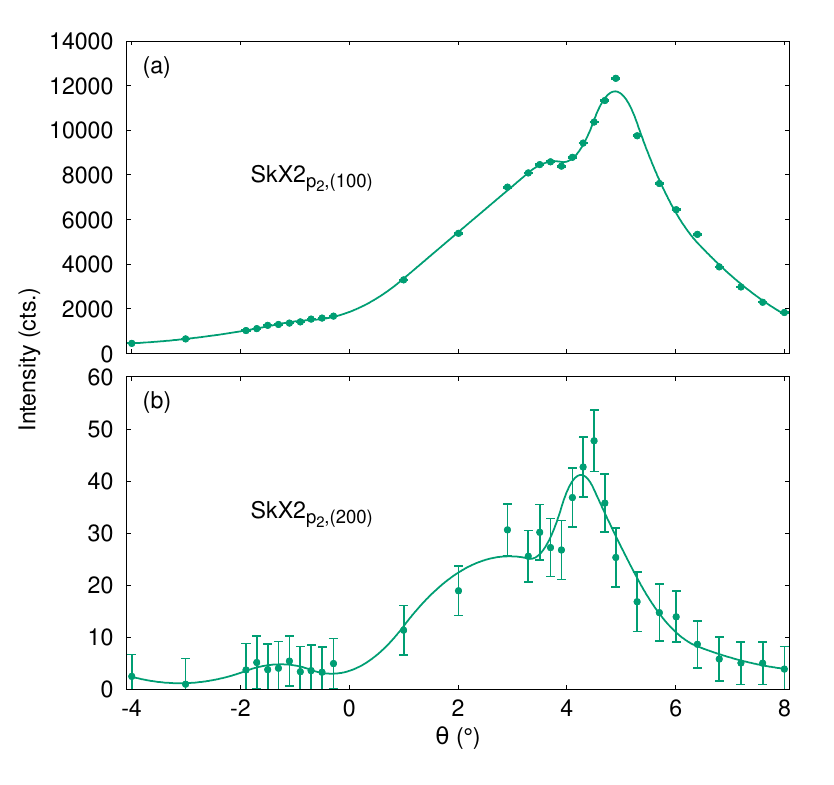}%
\caption{\label{fig:mosaicity_shift}
  Sample $\theta$-rotation scans around (a) the first-order \hkl(100) and (b) the higher-order \hkl(200) positions.
  The SkX2 phase was stabilized at $T=57$\,K and $\mu_0 H=17.35$\,mT using the FC protocol.  
  Intensity of the horizontal peak p2 was obtained by fitting the 2D intensity map to Eq.~\ref{eq:fit_2D}, and is plotted in the figure.
  Lines are guides for the eye.
}
\end{figure} 

The peak profile in the out-of-scattering-plane direction, obtained by the sample $\theta$-rotation scans, may provide additional support to distinguish the intrinsic higher-order contribution from the parasitic double scattering.
Specifically, for \hkl(200), due to the curvature of the Ewald sphere, the double scattering originates from a combination of two scattering processes with the first-order $\mathbf Q$-vectors of $(100) \pm (00\delta)$ [Fig.~\ref{fig:process}(a)].
Hence, the double reflection may appear at the identical $\theta$ as the first-order reflection, however, its peak width should be broadened by $2(2\theta) \sim 0.9^{\circ}$, as the double-scattering probability is a convolution of two single-scattering-probability functions shifted by $2\delta$.
In contrast, the intrinsic higher-order component should appear with almost the same width as that of the first-order reflection.

The $\theta$-scans were performed at a single selected $H-T$ point with long exposure time (at least 900~s at each $\theta$ angle) in order to attain decent statistics at the higher-order positions.
Scattering intensity at the first-order \hkl(100) and higher-order \hkl(200) positions (p2) was obtained using the 2D Gaussian fitting at each $\theta$ angle.
Figure~\ref{fig:mosaicity_shift} shows the resulting $\theta$ dependence of the reflection intensity.
Again, two-broad-peak feature is confirmed in the high-statistics data for the first-order peaks shown in Fig.~\ref{fig:mosaicity_shift}(a).
By comparing the peak position and profile of the first-order \hkl(100) and higher-order \hkl(200), one finds that peak profiles are mostly the same, indicating that the peak broadening expected for the double scattering does not occur.
In addition, the peak position shift is approximately $0.5^{\circ}$, which is in accordance with the expected shift given by the angle between $\mathbf{k}_{i,1}$ and $\mathbf{k}_{i,2}$, which is $2\theta \simeq 0.45^{\circ}$ for the first-order \hkl(100) peak.
This also supports that the intensity appearing at the higher-order positions are intrinsic, not arising from the parasitic double scattering.

\ifprintSections
\section{Monte Carlo Simulations}
\fi

To obtain additional support for the intrinsic origin of the observed higher-order reflections, and also to gain an insight into the deformation of the skyrmion texture related to the higher-order modulations, we have performed classical Monte Carlo (MC) simulations under a sweeping magnetic field.
For this purpose, a 2D square-lattice model with nearest neighbor ferromagnetic $J$ and antisymmetric DM interactions $D$ under external magnetic field $H_z$ was used in the MC simulation.
The ratio $D/J = d$ is fixed to $\tan(2\pi/13.75) \simeq 0.4917$, which corresponds to the skyrmion lattice constant $a_{\rm SkX} \simeq 16$, or to 64 skyrmions in the system.
Details of the MC simulation are given in Appendix~\ref{appendix:MC}.

\begin{figure}[h]
  \includegraphics[scale=0.4, trim=3cm 0cm 0cm 0cm]{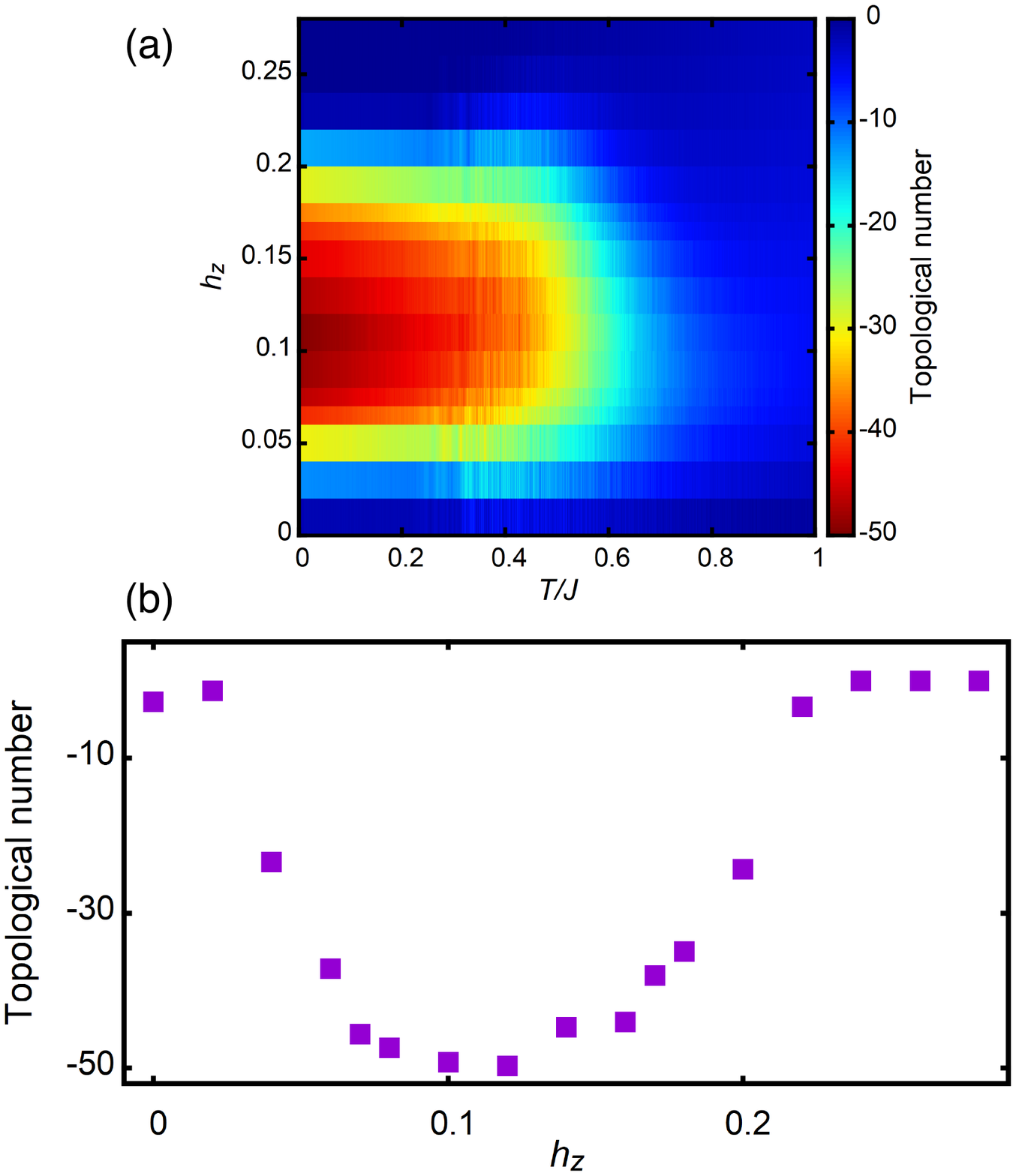}
  \caption{\label{mfig1} (a) Temperature and magnetic field dependence of the topological number $\Psi(T/J, h_z)$.
    The topological number is estimated from the thermally averaged spin configuration at each temperature during the simulated annealing under a fixed external magnetic field.
    For each magnetic field, the simulated annealing was performed from the paramagnetic temperature $T/J = 3$.
    (b) Magnetic field dependence of the topological number at $T/J = 0.01$.
}
\end{figure}

First, we performed simulated annealing runs from the paramagnetic $T/J = 3$ under several fixed external magnetic fields, to estimate the field range where the skyrmion-lattice phase is stabilized at low temperatures.
Figure~\ref{mfig1}(a) shows the topological number $\Phi$ evaluated from the thermally averaged spin configurations at each temperature obtained during the simulated annealing.
The magnetic field dependence of the topological number at $T/J = 0.01$ is shown in Fig.~\ref{mfig1}(b).
Finite topological numbers were observed at low temperatures in the magnetic field range of approximately $0.07 < h_z (\equiv H_z/J) < 0.16$, clearly indicating the formation of the skyrmion phase in this field range.

\begin{figure}[h]
  \includegraphics[scale=0.4, trim=3cm 0cm 0cm 0cm]{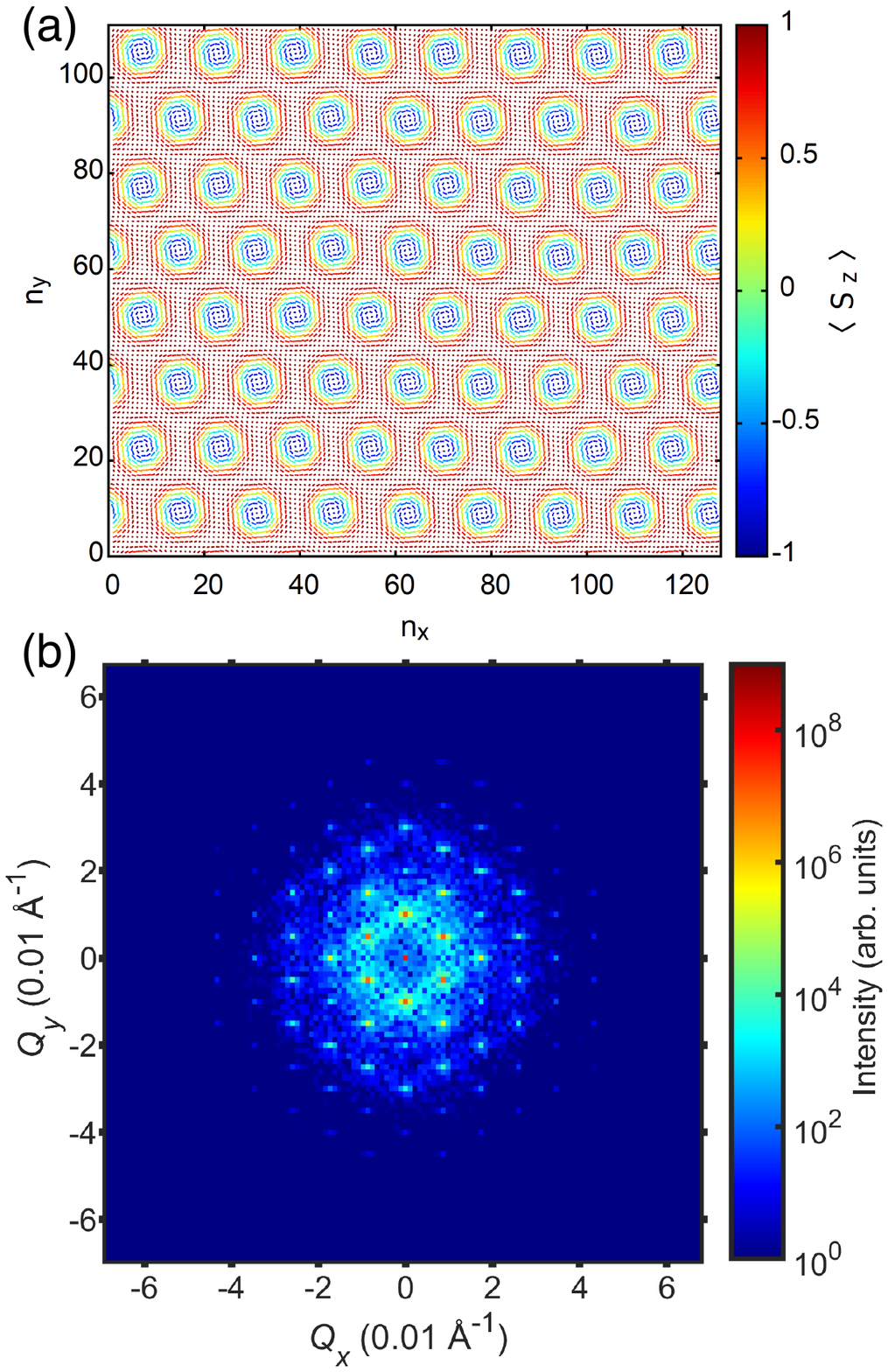}
  \caption{\label{mfig2} (a) Thermally averaged spin configuration obtained in the MC simulation at $T/J = 0.01$ and $h_z = 0.13$.
    The color indicates the $z$-component of the spin.
    (b) Scattering intensity map calculated from the thermally averaged spin configuration.
    $Q_x$ and $Q_y$ are scaled so that the first magnetic reflections coincide with the experimentally observed positions.
}
\end{figure}
\begin{figure}[h]
  \includegraphics[scale=0.28]{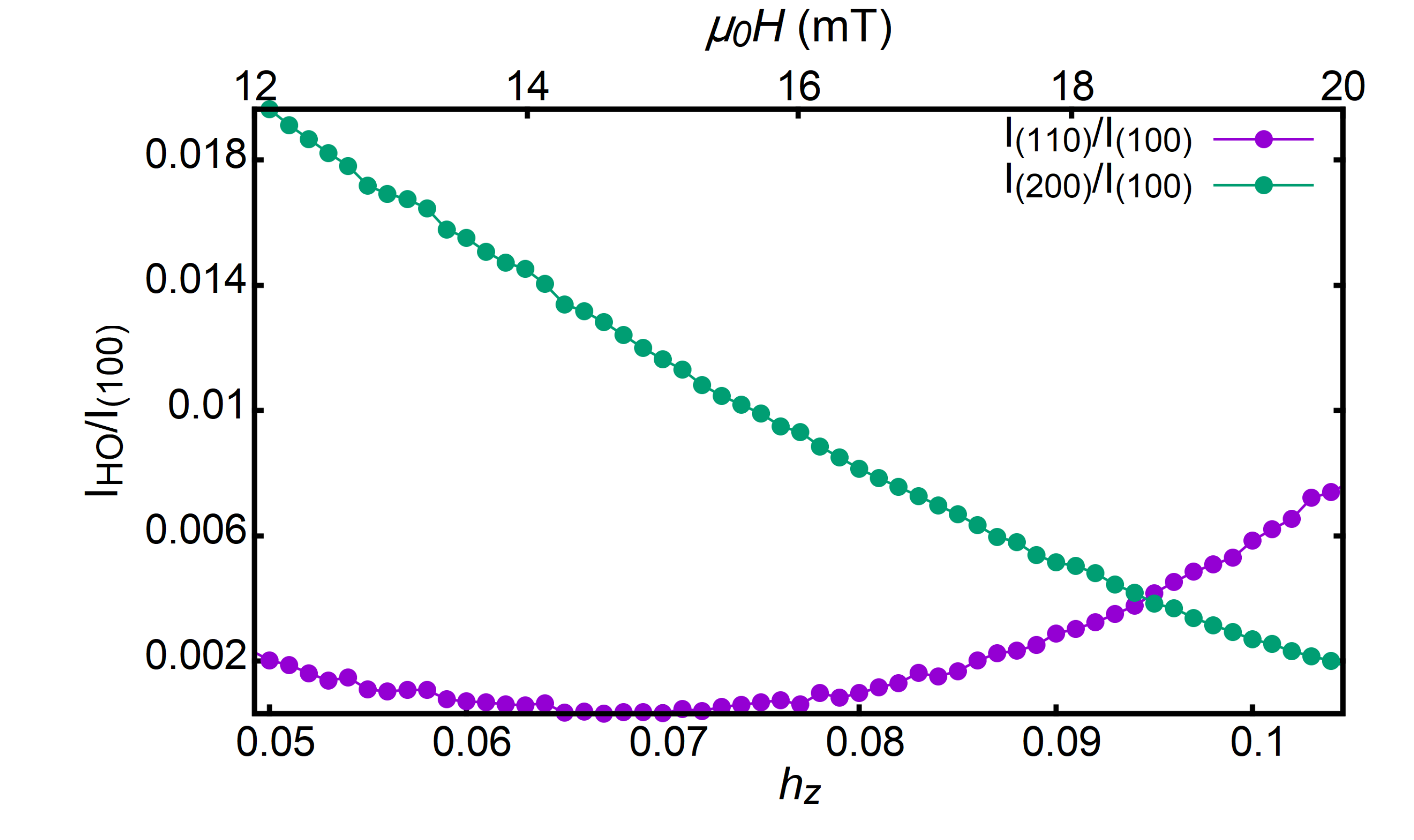}
  \caption{\label{mfig3} Magnetic field $h_z$ dependence of the higher-order reflection intensity for the 110 and 200 reflections.
    The magnetic field range corresponds to the region where the experimental observation of the higher-order reflection was performed [Fig.~\ref{fig:integ_intensity}(f)].
    Details of the result and estimation of the magnetic field range are given in the main text and Appendix~\ref{appendix:MC}.
  }
\end{figure}

A representative spin configuration at $T/J = 0.01$ and $h_z = 0.13$ is shown in Fig.~\ref{mfig2}(a), where a triangular-lattice structure of the magnetic skyrmions can be clearly confirmed.
Scattering intensity is then calculated using $I(\mathbf{Q}) = |\langle \mathbf{S}_{\perp}(\mathbf{Q}) \rangle|^2$, where $\langle \mathbf{S}_{\perp}(\mathbf{Q})\rangle = \langle \mathbf{S}(\mathbf{Q})\rangle - (\hat{\mathbf{Q}} \cdot \langle\mathbf{S}(\mathbf{Q})\rangle)\hat{\mathbf{Q}}$ and $\langle \mathbf{S}(\mathbf{Q})\rangle = \sum_{\mathbf{R}}\langle\mathbf{S}_{\mathbf{R}}\rangle \exp(-i \mathbf{Q}\cdot\mathbf{R})$.
The resulting scattering pattern is shown in Fig.~\ref{mfig2}(b).
Note that the intensity is shown on a logarithmic scale.
In addition to the six-fold first-order peaks, weak but clear higher-order reflections can be seen in the figure.

Once the stabilization has been reached at $T/J = 0.01$ (as above), the magnetic field is either slowly increased or decreased in the MC simulation to destabilize the skyrmion-lattice phase while monitoring the higher-order-reflection intensity.
The magnetic-field dependence for the representative reflections 110 and 200 in the field range corresponding to the experimentally observed region [Fig.~\ref{fig:integ_intensity}(f)] is plotted in Fig.~\ref{mfig3}.
It can be clearly seen that $I_{(110)}/I_{(100)}$ increases in the field range of $0.07 < h_z < 0.16$, whereas $I_{(200)}/I_{(100)}$ decreases.
This contrasting behavior for the two higher-order reflections is qualitatively consistent with the experimental observation shown in Fig.~6(f).
One small discrepancy may be found at the lower field edge ($h_z = 0.07$ or $\mu_0 H = 13.5$~mT
), where the intensity decreases for both the reflections in the experiment, while the $I_{(200)}/I_{(100)}$ has still a large value in the MC simulation.
We think that this is likely due to the proximity to the phase boundary, as discussed in Appendix~\ref{appendix:MC}.
Here, we suggest that the overall qualitative correspondence of the field dependence between the experiment and MC simulation further supports the intrinsic origin of the higher-order reflections observed in our SANS experiment.

\ifprintSections
\section{Discussion}
\fi
In the present work, we found that in \Cu the parasitic double scattering is rather weak, and the dominant contribution stems from the intrinsic higher-order modulation of the skyrmion lattice structure.
When considering the intensity observed at the higher-order-peak positions as such, both higher-order-peak sets show a mostly constant ratio $I_{\textrm{ho}}/I_{(100)}$ under temperature variation, indicating that the structure of the skyrmion lattice remains mostly the same at all temperatures.
The slight decrease close to the boundaries hints at a more regular structure.
In contrast, the magnetic field dependence of the higher-order scattering differs for \hkl(110) and \hkl(200).
While the ratio $I_{(200)}/I_{(100)}$ decreases at low and high magnetic fields, $I_{(110)}/I_{(100)}$ increases at higher fields.
The different $H$-dependence for the ratios $I_{(110)}/I_{(100)}$ and $I_{(200)}/I_{(100)}$ suggests that the skyrmions are not circular, and its distortion, induced by the external magnetic field, depends on external magnetic fields.
Details of the distortion are discussed in Appendix~\ref{appendix:MC}.

It should be noted that the magnetic field dependence of the intrinsic \hkl(110) contribution has been investigated in an earlier MnSi study~\cite{Adams2011}, indicating monotonically increasing behavior as the magnetic field is increased.
This is indeed qualitatively the same as the present observation.
It was pointed out that the intrinsic \hkl(110) contribution is strongly suppressed at the lower field boundary ($B_{\rm int} \simeq 170$~mT) in MnSi, which is a key property ensuring the topological nature of the skyrmion-lattice structure.
Also in the presently studied \Cu, we found that the \hkl(110) intensity shows its minimum in the vicinity of the lower-$\mu_0 H$ phase boundary (Fig.~\ref{fig:integ_intensity}), confirming the topological nature of its skyrmion-lattice phases.
On the other hand, the \hkl(200) intensity shows its minimum at the higher-$\mu_0 H$ boundary.
The MC simulation using the simple Hamiltonian Eq.~\ref{eq:modelH} qualitatively reproduces the observed $H$-dependence for both the \hkl(110) and \hkl(200) intensities (Fig.~\ref{mfig3} or Fig.~\ref{appfig1}), suggesting that the higher-order modulations have a thermodynamic origin, reflecting a different $\mathbf{Q}$-dependence of free-energy surface at the \hkl(110) and \hkl(200) positions~\cite{}, and not related to the single-ion anisotropy effect, which is the more typical origin of ``squaring-up'' distortion~\cite{Izyumov84,Sato94}.
It is indeed shown in Appendix B that the distortion of the skyrmions by the external magnetic field is driven by the formation of new swirlings with the superlattices characterized by the higher-order modulations \hkl(110) and \hkl(200) (see Fig.~\ref{appfig2}).

It may be noteworthy that in the earlier work on MnSi, a significant amount of double scattering contribution was observed.
Since the SANS experiment setting (such as wavelength and resolution) is similar in the earlier MnSi and present \Cu experiment, and the $|\mathbf Q|$ of the first-order reflection is much smaller than that in the MnSi experiment, the double-scattering contribution should in principle appear more prominently in the present experiment.
Nonetheless, our experiment showed that the double scattering component is relatively weak compared to the intrinsic higher-order modulation.
This may be because the higher-order modulation is more significant in the present $T$ and $H$ range in \Cu.
Indeed, the higher-order intensity is roughly $2 \times 10^{-3}$ of the first-order peak, as inferred in Fig.~\ref{fig:integ_intensity}(c) or Fig.~\ref{fig:integ_intensity}(f), which is larger than the higher-order contribution in MnSi~\cite{Adams2011}.
Another possibility is the suppression of the double reflection due to the accidental misorientation of the sample.
Although the six-fold first-order peaks are simultaneously observed in the present setup, we know that the sample $[\bar{1} 1 0]$ axis is tilted from the vertical axis by approximately 5 degrees.
Hence, the double scattering condition for \hkl(110) is far from optimal, and consequently, the double scattering intensity may be suppressed in comparison to what is achievable with the Renninger scans.

%
%
%


\ifprintSections
\section{Conclusion}
\fi
SANS measurements were performed in \Cu to elucidate the origin of the higher-order scattering.
We found that significant higher-order scattering appears in the SANS patterns for both the SkX1 and SkX2 phases stabilized under certain $H-T$ history.
Notably, however, the higher-order intensity at the $\mathbf{Q}$ position combining modulations from the two phases, such as $\mathbf{Q} = \mathbf{Q}_{\rm SkX1} + \mathbf{Q}_{\rm SkX2}$, does not appear.
This absence clearly indicates that the parasitic double scattering is much weaker than the intrinsic higher-order modulations.
The temperature and magnetic-field dependence of the higher-order intensity, as well as its peak profile along the out-of-scattering-plane direction, supports the weaker contribution of the parasitic double scattering.
It is further found that the two higher-order modulations, \hkl(110) and \hkl(200), show contrasting magnetic-field dependence; the former increases as $H$ is increased, whereas the latter decreases.
This clearly indicates that, in Cu$_2$OSeO$_3$, skyrmions are weakly distorted, and the distortion is field-dependent in a way that the dominant higher-order modulation switches from \hkl(110) in the low-field region to \hkl(200) in the high-field region.
The MC simulation using the simple two-dimensional square lattice with nearest neighbor ferromagnetic and Dzyaloshinskii-Moriya interactions qualitatively reproduces the observed field-dependence of the higher-order modulations.
The MC simulation further suggests that the higher-order modulations correspond to the superlattices of weak swirlings appearing in the middle of the original triangular-latticed skyrmions.

\vspace{0.1cm}

\begin{acknowledgments}
The authors thank N. Nagaosa, H. Ronnow for stimulating discussions, and S. Yunoki and S. Zhang for providing us unpublished results of a large-scale Monte-Carlo simulation.
This work was partly supported by Grants-In-Aid for Scientific Research (24224009, 19H01834, 19K21839, 19H05824, 21H04440, 21H04990, 21K18595) from MEXT of Japan, PRESTO (grant no. JPMJPR18L5) from JST, and Asahi Glass Foundation.
Travel expense for the experiment was partly sponsored by the General User Program of ISSP-NSL, University of Tokyo.
Work at IMRAM was partly supported by the Research Program ``Dynamic Alliance for Open Innovation Bridging Human, Environment and Materials''.
JDR is an International Research Fellow  of the  Japan Society for the Promotion of Science.
\end{acknowledgments}

\appendix

\section{Intensity variation in sample $\theta$-rotation scans}\label{appendix:rotationscans}

\begin{figure*}[ht]
\includegraphics{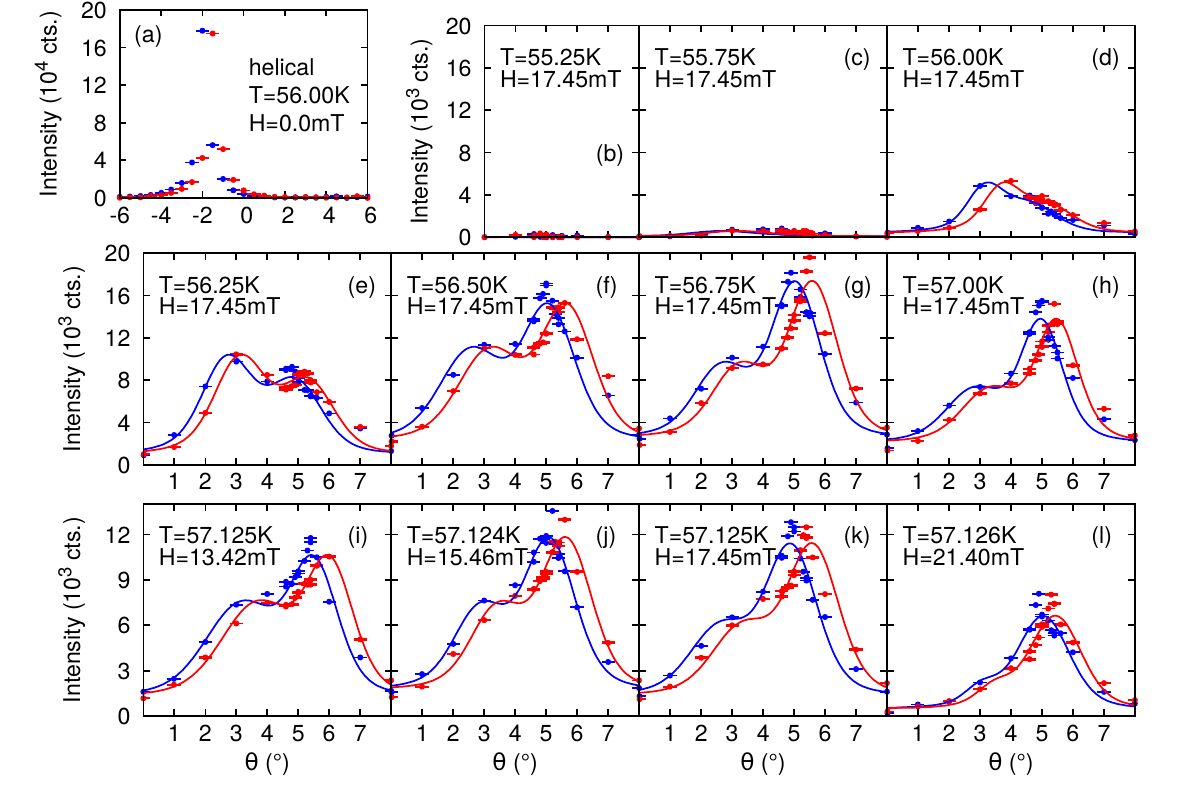}%
\caption{\label{fig:sup_mosaicity}
  (a) Result of the sample $\theta$-rotation scan for the helical peaks, used in the fitting of skyrmion peaks as an instrumental resolution function for the $\theta$-direction.
  (b-l) Results of the sample $\theta$-rotation scans for the first-order peaks on the horizontal axis [p2 (blue) and p5 (red)].
  The skyrmion lattice at each temperature/magnetic-field is stabilized using the FC protocol.
  The observed intensity (points) is compared with the fit (line) using Eq. \ref{eq:fit_mosaicity}.
  The temperature dependence is displayed in (b-h) and the magnetic field dependence in (i-l).
}
\end{figure*}

\begin{figure}[h]
\includegraphics{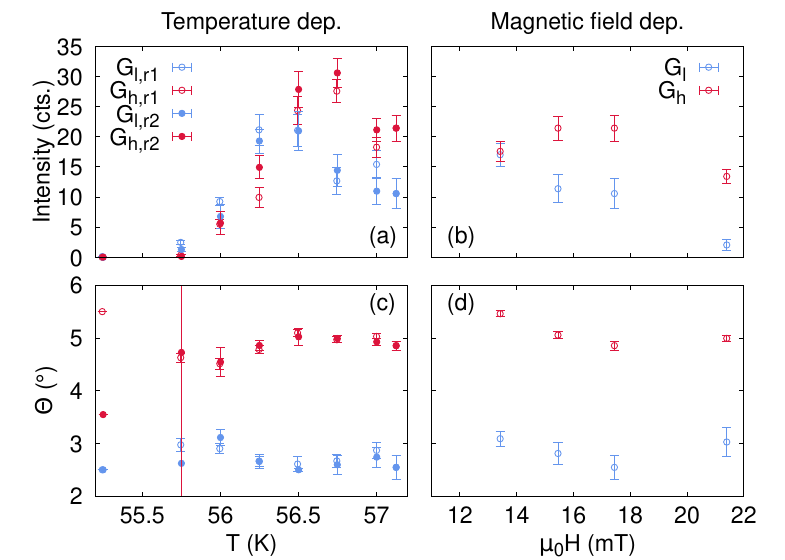}%
\caption{\label{fig:mosaicity_fitting_appendix}
  Parameters obtained from the two 1D-Gaussian fitting of the sample $\theta$-rotation scans shown in Fig.~\ref{fig:sup_mosaicity}.
  $G_l$ and $G_h$ stand for the Gaussian-shaped peaks appearing at lower and higher $\theta$ angles, respectively.
  Temperature dependence of the parameters obtained from the scans at $\mu_0 H=17.45$\,mT is shown in (a) and (c), whereas the external magnetic-field dependence at $T = 57.125$~K is shown in (b) and (d).
  The temperature dependence was measured twice [run 1 (r1) and 2 (r2)] and shows a reasonable reproducibility.
}
\end{figure}

Results of the sample $\theta$-rotation scans for the first order \hkl(100) peak are shown in Fig.~\ref{fig:sup_mosaicity}.
The measurements were performed at various temperatures and magnetic fields.
It can be clearly seen that relative intensity and position of the two peaks depends on magnetic field and temperature.
On the other hand, the overall shift between the curves for the p2 and p5 reflections is always $\delta_5$, independent of temperature and magnetic field, being consistent with the geometrical condition for the Bragg reflection taking account of the curvature of the Ewald sphere.
Applying the model described in Subsection \ref{subsec:structure}, we obtained reasonable approximate analytic profile function at all temperatures and magnetic fields (see solid lines in Fig.~\ref{fig:mosaicity_fitting} and Fig. \ref{fig:sup_mosaicity}).
The optimal fitting parameters, peak intensity and position, are shown in the Fig.~\ref{fig:mosaicity_fitting_appendix}.
It may be noteworthy that the positions $\theta_0$ of the two Gaussians in the intensity variations show an antagonistic temperature dependence, while with the magnetic field they change seemingly in accordance.
In addition, the dependence of the intensity on temperature and magnetic field differs for the two Gaussians.
At low temperatures SkX2$^\prime$ dominates while at high temperatures and higher fields SkX2$^{\prime\prime}$ increases.
We will not go into detailed analysis, but just note here that an intricate mechanism related to the sample shape may be necessary to understand this behavior.

\section{Details of Monte Carlo simulations}\label{appendix:MC}

The classical MC simulation was performed using the standard Metropolis algorithm combined with the simulated annealing method.
The success rate of the spin flip at each temperature is fixed to 50~\%~\cite{Creutz1987} using the Gaussian move algorithm to enhance the efficiency for the thermal stabilization~\cite{EvansRFL2014,AlzateCardona19}.
The system size was chosen as $(N_x, N_y) = (127, 110)$; this particular rectangular shape was chosen to reduce the dislocations introduced by the mismatch between the periodic boundary condition for the triangular lattice and the system dimension~\cite{NishikawaY19}.

\begin{figure}[h]
  \includegraphics[scale=0.28]{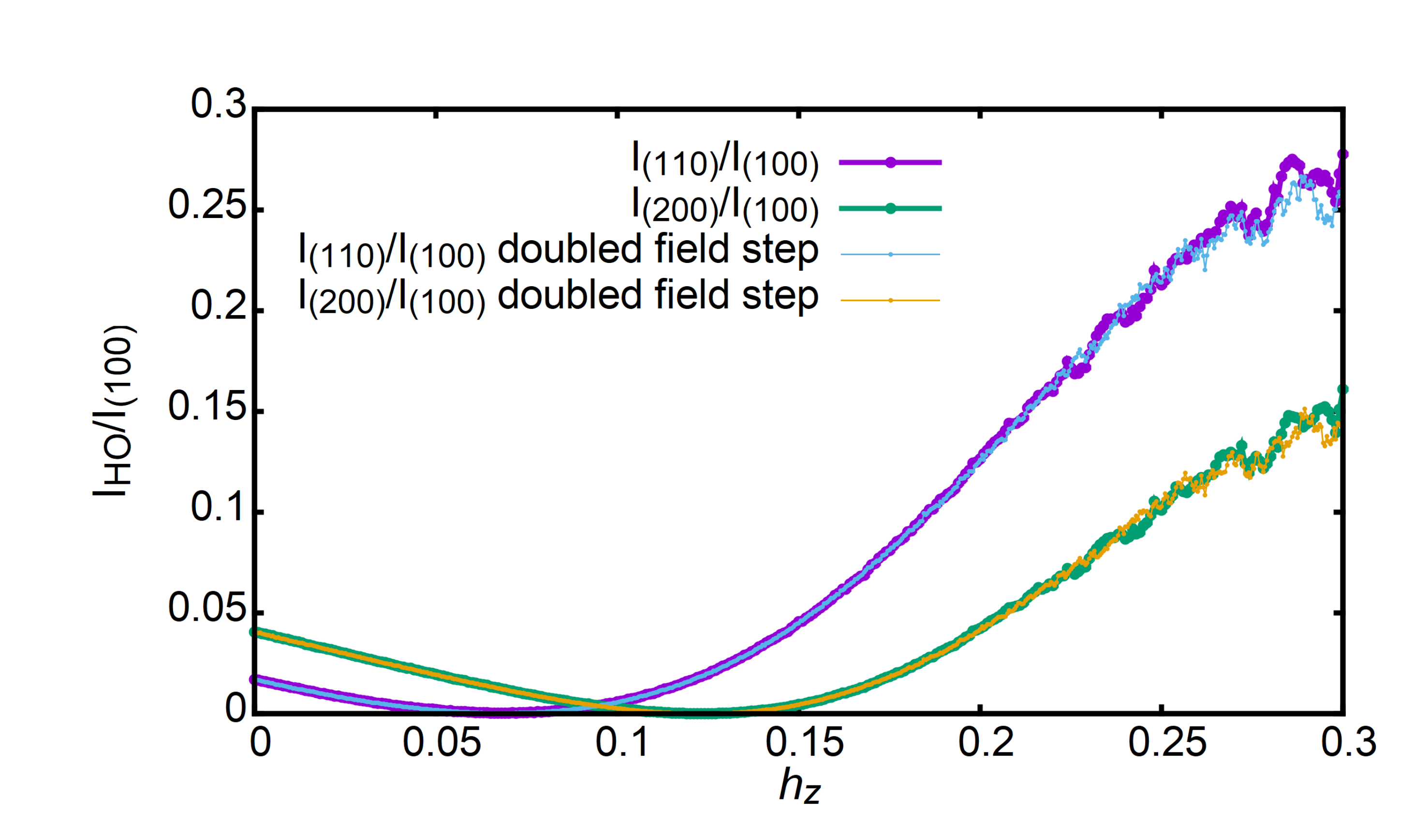}
  \caption{\label{appfig1} Magnetic field $h_z$ dependence of the higher-order reflection intensity for the 110 and 200 reflections in the wide field range $0 < h_z < 0.3$.
    Details of the result and estimation of the magnetic field range are given in the main text and Appendix.
  }
\end{figure}

\begin{figure*}[h]
  \includegraphics[scale=0.6, trim=0cm 0cm 0cm 3cm]{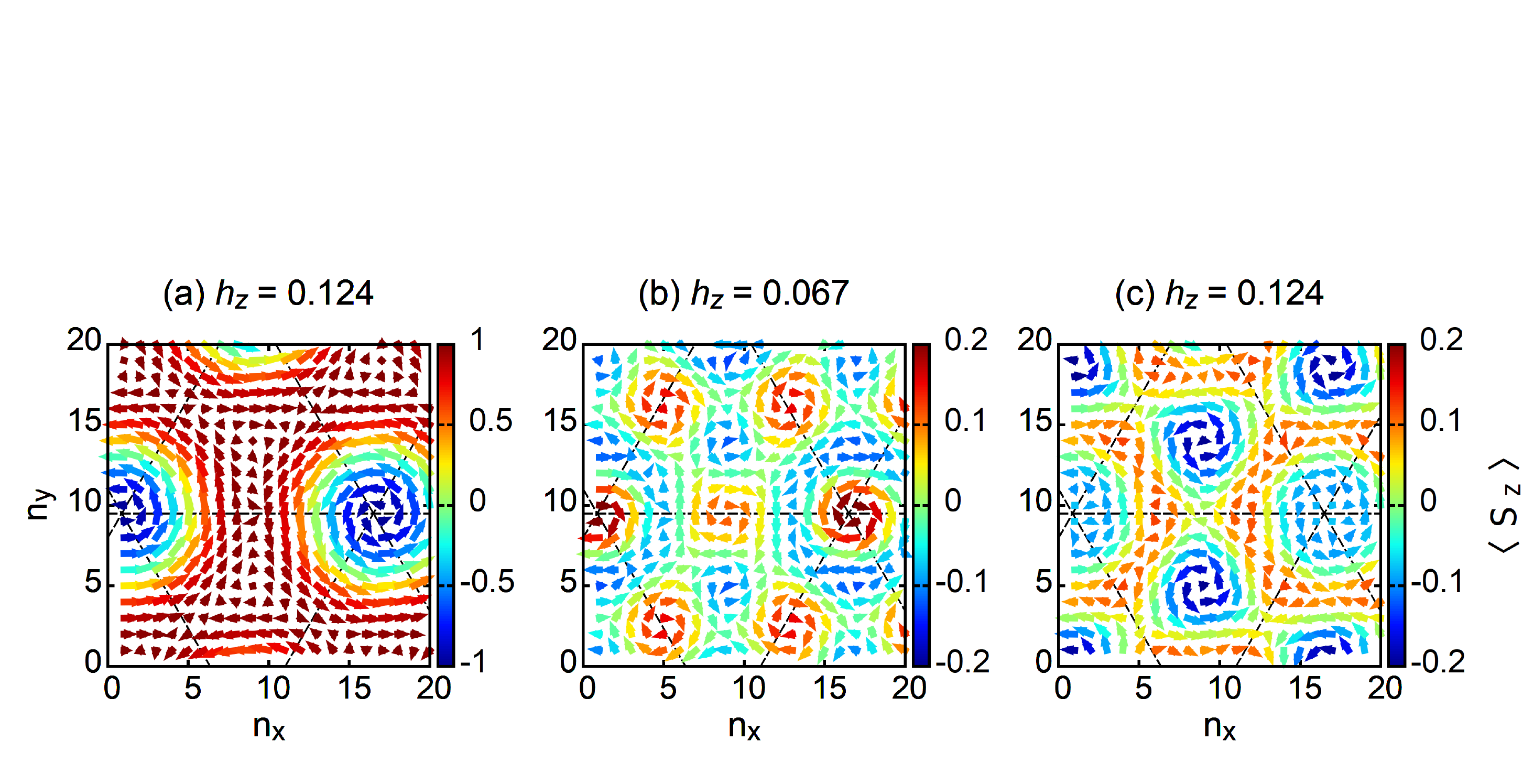}
  \caption{\label{appfig2}
    (a) Magnified plot of skyrmion lattice formed at $T/J = 10^{-8}$ and $h_z = 0.124$.
    (b,c) Spin configuration corresponding to the higher-order modulations obtained by inverse Fourier transform of $\mathbf{S}(\mathbf{Q})$ with removing the first-order reflections ($|\mathbf{Q}| < 0.0166$~\AA$^{-1}$.)
    (b) Spin configuration at $h_z = 0.067$ was used, where the dominant higher-order modulation is \hkl(200).
    In addition to the original skyrmion centers, other swirling appears at the center of the triangular bonds.
    (c) Spin configuration at $h_z = 0.124$ was used, where the dominant higher-order modulation is \hkl(110).
    In addition to the original skyrmion centers, other swirling appears at the center of the triangular unit cell.
  }
\end{figure*}

In the MC simulation, we used the following well-established spin Hamiltonian consisting of the nearest-neighbor ferromagnetic and antisymmetric DM interactions on the 2D square lattice~\cite{YiSD09}:
\begin{widetext}
\begin{equation}\label{eq:modelH}
  {\cal H}/J = -\sum_{\mathbf{R}}\mathbf{S}_{\mathbf{R}}\cdot(\mathbf{S}_{\mathbf{R}+\hat{\mathbf{x}}} + \mathbf{S}_{\mathbf{R}+\hat{\mathbf{y}}})
  -d\sum_{\mathbf{R}}[(\mathbf{S}_{\mathbf{R}} \times \mathbf{S}_{\mathbf{R}+\hat{\mathbf{x}}}) \cdot \hat{\mathbf{x}}
  + (\mathbf{S}_{\mathbf{R}} \times \mathbf{S}_{\mathbf{R}+\hat{\mathbf{y}}}) \cdot \hat{\mathbf{y}}]
  - h_z \sum_{\mathbf{R}}S_{\mathbf{R}}^z,
\end{equation}

\end{widetext}
where $d = D/J$ and $h_z = H_z /J$ are the dimensionless DM-interaction and external-field (along $z$) parameters, respectively.
We used $d = \tan(2\pi/13.75) \simeq 0.4917$, which corresponds to the skyrmion lattice constant $a_{\rm SkX} \simeq 16$.

For the phase diagram study shown in Fig.~\ref{mfig1}(a), at each fixed magnetic field, the simulated annealing was started from sufficiently high temperature $T/J = 3$, down to the base temperature $T/J = 0.0001$.
The temperature was decreased linearly in 200 temperature steps for $3 > T/J \geq 1.0$, 1400 steps for $1 > T/J \geq 0.01$, 100 steps for $0.01 > T/J \geq  0.001$, and 100 steps for $0.001 > T/J \geq 0.0001$.
At each temperature we waited 7000 MCS for thermally equilibration.
In the simulated annealing runs, thermal averaged spin configuration is evaluated using the configurations in the last $3500$ Monte Carlo steps (MCS) at each temperature.
The topological number $\Psi(T/J, h_z)$ shown in Fig.~\ref{mfig1}(b) is estimated from the thermally averaged configurations using the following equation:
\begin{equation}
  \begin{split}
    &\Psi(T/J, h_z) = \frac{1}{4\pi}\sum_{\mathbf{R}}\frac{\langle\mathbf{S}_{\mathbf{R}}\rangle}{|\langle \mathbf{S}_{\mathbf{R}}\rangle|}\cdot \left [ \left ( \frac{\langle\mathbf{S}_{\mathbf{R}+\hat{\mathbf{x}}}\rangle}{|\langle\mathbf{S}_{\mathbf{R}+\hat{\mathbf{x}}}\rangle |} -\frac{\langle\mathbf{S}_{\mathbf{R}}\rangle}{|\langle\mathbf{S}_{\mathbf{R}}\rangle|}\right ) \right. \\
      & \left . \times \left ( \frac{\langle \mathbf{S}_{\mathbf{R}+\hat{\mathbf{y}}}\rangle }{|\langle \mathbf{S}_{\mathbf{R}+\hat{\mathbf{y}}}\rangle |}-\frac{\langle \mathbf{S}_{\mathbf{R}}\rangle }{|\langle \mathbf{S}_{\mathbf{R}}\rangle |} \right ) \right ].
  \end{split} 
\end{equation}
It should be noted that since the system is two dimensional, well-defined ordering does not exist at finite temperature.
Nonetheless, clear formation of the skyrmion phase is seen in Fig.~\ref{mfig1}, indicating that the skyrmion phase has sufficiently slow dynamics at low temperatures.

In the MC simulation, the skyrmion-lattice phase was observed in the magnetic field range of $0.07 < h_z < 0.16$ (Fig.~\ref{mfig1}).
Assuming that the range where the skyrmion-lattice phase is found in MC corresponds to the experimentally observed range ($15 \lesssim \mu_0 H \lesssim 28$~mT \cite{Makino2017}), we estimate that the field range of the present higher-order experiment ($13.5 < \mu_0 H < 19.5$~mT, see Fig.~\ref{fig:integ_intensity}(f)) roughly corresponds to $0.07 < h_z < 0.1$.
This is the range where the MC result is shown in Fig.~\ref{mfig3}.

The field sweep simulation was performed in two steps.
First, we performed the MC simulated annealing from the paramagnetic temperature to $T/J = 0.01$ under $h_z = 0.13$.
After confirming the formation of the skyrmion-lattice, we swept the magnetic field either to $h_z = 0$ with 130 field steps, or to $h_z = 0.3$ with 170 steps.
At each field, we waited 7000 MCS for equilibration.
The ratio of the two higher-order intensities $I_{(110)}$ and $I_{(200)}$ to the first-order intensity $I_{(100)}$ is shown in Fig.~\ref{appfig1} in a wide field range $0 < h_z < 0.3$.
We note that sweep rate does not significantly modify the results; we checked the sweep rate dependence by performing the field sweep with a doubling of field steps with the same waiting MCS at each field (260 steps for $h_z \rightarrow 0$, while 340 steps for $h_z \rightarrow 0.3$), however, no significant difference between the two results was observed.
(Compare the thicker and thinner lines in Fig.~\ref{appfig1}.)
On the other hand, the higher-order reflections, as well as main skyrmion-lattice reflections, survive even outside the skyrmion-formation range $0.07 < h_z < 0.16$.
This reflects the fact that the skyrmion-lattice phase is indeed quite robust to its destruction to the trivial helical or ferromagnetic phase.
It may be further noted that the lowest magnetic field in Fig.~\ref{mfig3} ($h_z = 0.07$
 corresponding to the experimental $\mu_0 H = 13.5$~mT) is indeed slightly outside the skyrmion-formation range.
Therefore, although the MC-obtained $I(200)/I(100)$ increases as $h_z$ decreases to 0.07 as shown in Fig.~\ref{mfig3} or Fig.~\ref{appfig1}, we think the instability of the skyrmion-lattice phase itself would result in the decrease of the higher-order reflection intensity in reality.

Distortion of skyrmions under the magnetic field is further checked by performing the inverse Fourier transform.
As a reference, Fig.~\ref{appfig2}(a) shows the averaged spin configuration obtained at the base temperature $T/J = 10^{-8}$ obtained under $h_z = 0.124$.
To increase visibility, only a small area ($20 \times 20$) is depicted.
The triangular lattice of skyrmions in this figure corresponds to the first-order reflections in the SANS patterns.
This real space spin arrangement was firstly Fourier transformed to obtain $\langle \mathbf{S}(\mathbf{Q}) \rangle$, and then the $\langle \mathbf{S}(\mathbf{Q})\rangle$ with $|\mathbf{Q}| \geq 0.0166$~\AA$^{-1}$ is inverse Fourier transformed again into real space.
This way, the contribution of the first-order modulation (appearing at $|\mathbf{Q}| \simeq 0.01$~\AA$^{-1}$) is removed, and consequently, the spin modulation originating only from the higher-order modulation may be depicted.
Fig.~\ref{appfig2}(b) and \ref{appfig2}(c) shows thus obtained real-space spin configurations for $h_z = 0.067$ and $h_z = 0.124$.
In the former case, the dominant higher-order modulation is \hkl(200), whereas \hkl(110) is dominant for the latter.
It can be clearly seen that in addition to the swirlings appearing at the original triangular-lattice vertices, those at the bond centers (former) or at the unit-cell centers (latter) appear.
It may be noteworthy that the helicity~\cite{GungorduU2016} of the higher-order swirlings are opposite; those for the \hkl(110) have the same helicity to the first-order skyrmions, whereas opposite helicity is seen for \hkl(200).

\bibliography{higher_order.bib}

\end{document}



\title{Supplemental information for higher-order modulations in the skyrmion-lattice phase of \Cu}


\author{Johannes D. Reim}
\email{johannes.reim@rwth-aachen.de}
\author{Shinnosuke Matsuzaka}
\email{shinnosuke.matsuzaka.r2@dc.tohoku.ac.jp}
\author{Koya Makino}
\author{Seno Aji}
\author{Ryo Murasaki}
\author{Daiki Higashi}
\author{Daisuke Okuyama}
\affiliation{Institute of Multidisciplinary Research for Advanced Materials,
Tohoku University, 2-1-1 Katahira, Sendai 980-8577, Japan}

\author{Yusuke Nambu}
\affiliation{Institute for Materials Research, 
Tohoku University, 2-1-1 Katahira, Sendai 980-8577, Japan}
\affiliation{Organization for Advanced Studies, 
Tohoku University, 2-1-1 Katahira, Sendai 980-8577, Japan}
\affiliation{FOREST, Japan Science and Technology Agency,
Kawaguchi, Saitama 332-0012, Japan}

\author{Elliot P. Gilbert}
\author{Norman Booth}
\affiliation{Australian Centre for Neutron Scattering, Australian Nuclear Science and Technology Organization,
Kirrawee DC, New South Wales 2232, Australia}

\author{Shinichiro Seki}
\affiliation{RIKEN Center for Emergent Matter Science (CEMS), Wako, Saitama 351-0198, Japan}
\affiliation{Department of Applied Physics and Quantum Phase Electronics Center (QPEC), University of Tokyo, Tokyo 113-8656, Japan}

\author{Yoshinori Tokura}
\affiliation{RIKEN Center for Emergent Matter Science (CEMS), Wako, Saitama 351-0198, Japan}
\affiliation{Department of Applied Physics and Quantum Phase Electronics Center (QPEC), University of Tokyo, Tokyo 113-8656, Japan}

\author{Taku J Sato}
\affiliation{Institute of Multidisciplinary Research for Advanced Materials,
Tohoku University, 2-1-1 Katahira, Sendai 980-8577, Japan}


\date{\today}

\maketitle

\onecolumngrid
\section{Magnetic-field dependence of higher-order scattering intensity}

\begin{figure}[h!]
\includegraphics{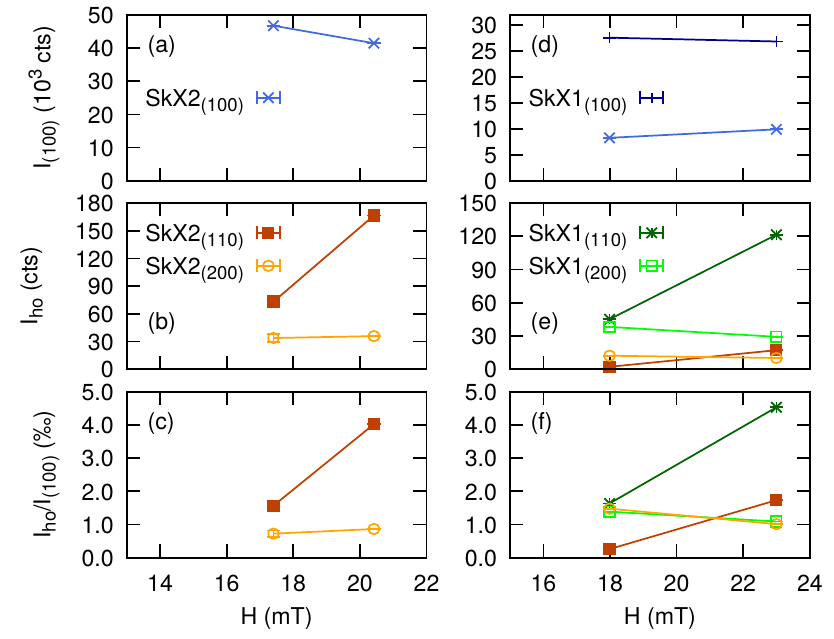}%
\caption{\label{fig:sup_fielddep}
  Magnetic field dependence of the intensity at the higher order peak positions.
  (a-c) Results for the SkX2 phase stabilized at $T = 57.0$~K using the FC protocol and (d-f) those for the coexisting SkX1 and SkX2 phases stabilized at $T = 56.5$~K using the FW protocol are shown. 
}
\end{figure}

Additional datasets for the magnetic-field dependence of the higher-order intensity is obtained at $T = 57.0$~K and $T = 56.5$~K are shown in Figs.~\ref{fig:sup_fielddep}(a-c) and \ref{fig:sup_fielddep}(d-f), respectively.
The former dataset was taken for the SkX2 phase stabilized using the FC protocol, whereas the latter was for the coexisting SkX1 and SkX2 phases stabilized using the FW protocol.
The \hkl(110) intensity shows clear increasing behavior for increasing field, whereas weakly decreasing [Fig.~\ref{fig:sup_fielddep}(b)] or almost flat [Fig.~\ref{fig:sup_fielddep}(a)] behavior may be seen for the \hkl(200) reflection.
These results indicates that the magnetic field dependence of the higher-order intensity does not depend on the temperature-field history under which the skyrmion-lattice phase(s) is (are) stabilized.

\section{Long exposure SANS patterns used for the higher-order scattering estimation}
To obtain the temperature and magnetic-field dependence of the higher-order scattering intensity, a number of long-exposure SANS patterns were obtained.
The representative two SANS intensity maps were shown in the main text as Figs.~2 and 3.
Below, all the long-exposure SANS patterns used to obtain the higher-order scattering intensity (shown in Fig.~6) are summarized.
The sample $\theta$-rotation was fixed to the maximum-intensity position, i.e. $\theta \simeq 5.1^{\circ}$.

%
%
%
%
%
%
%
%
%
%
%

%

%
\begin{figure}[h]
\includegraphics{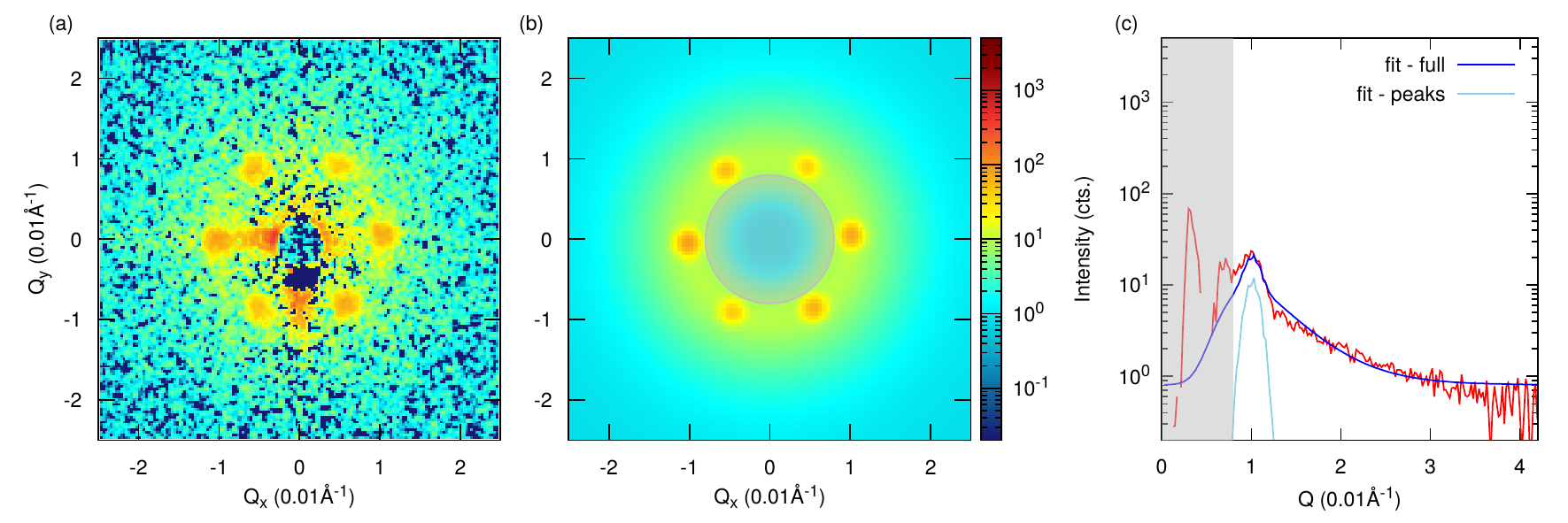}%
\caption{\label{fig:sup_map_94736}
  2D SANS intensity map obtained at $T=55.25$\,K and $\mu_0 H=17.4$\,mT using FC protocol.
}
\end{figure}

\begin{figure}[h]
\includegraphics{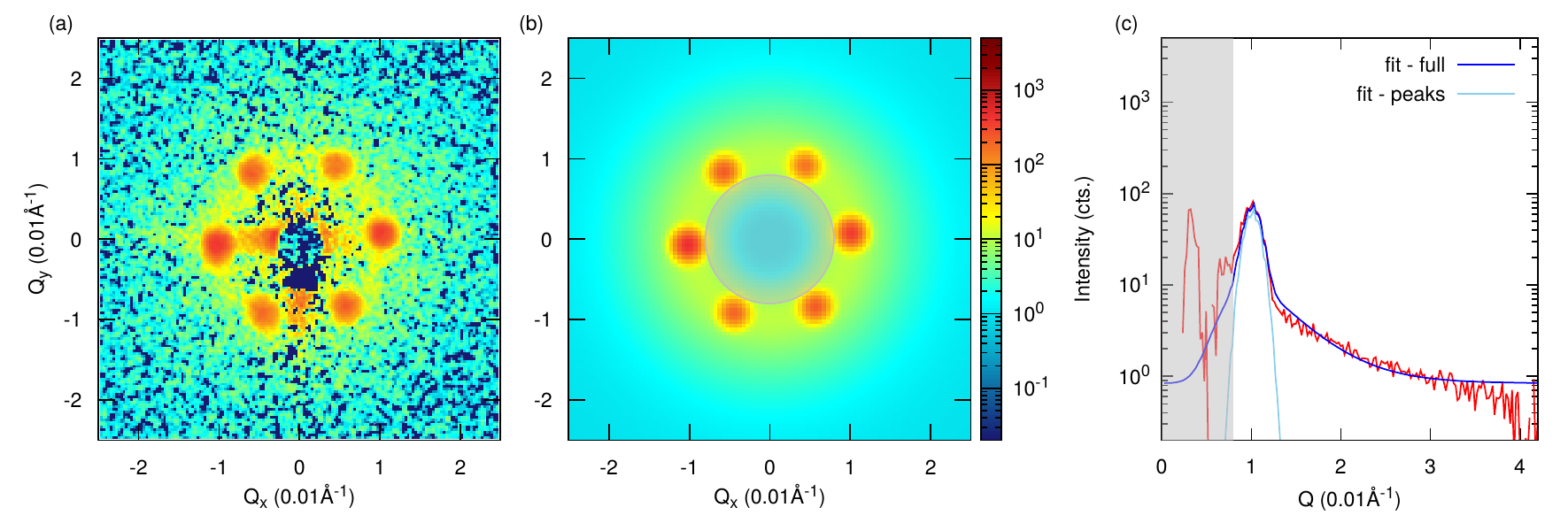}%
\caption{\label{fig:sup_map_94706}
  2D SANS intensity map obtained at $T=55.75$\,K and $\mu_0 H=17.4$\,mT using FC protocol.
}
\end{figure}

%

%

%
\begin{figure}
\includegraphics{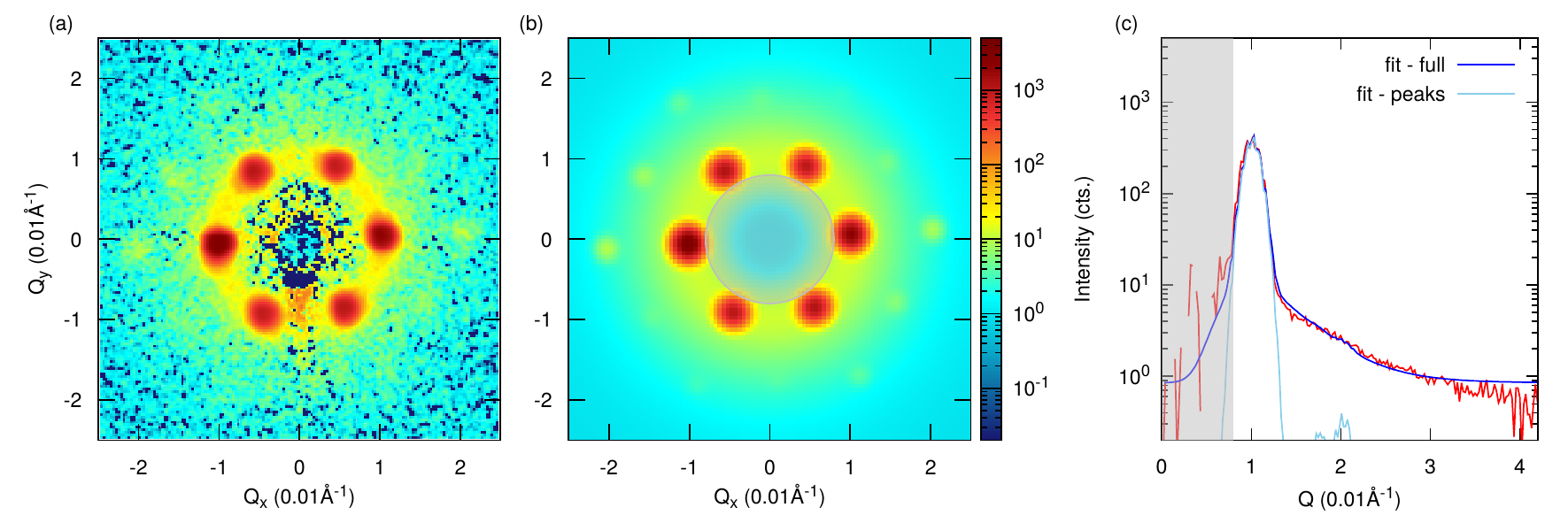}%
\caption{\label{fig:sup_map_95469}
  2D SANS intensity map obtained at $T=56.00$\,K and $\mu_0 H=17.4$\,mT using FC protocol.
}
\end{figure}

\begin{figure}
\includegraphics{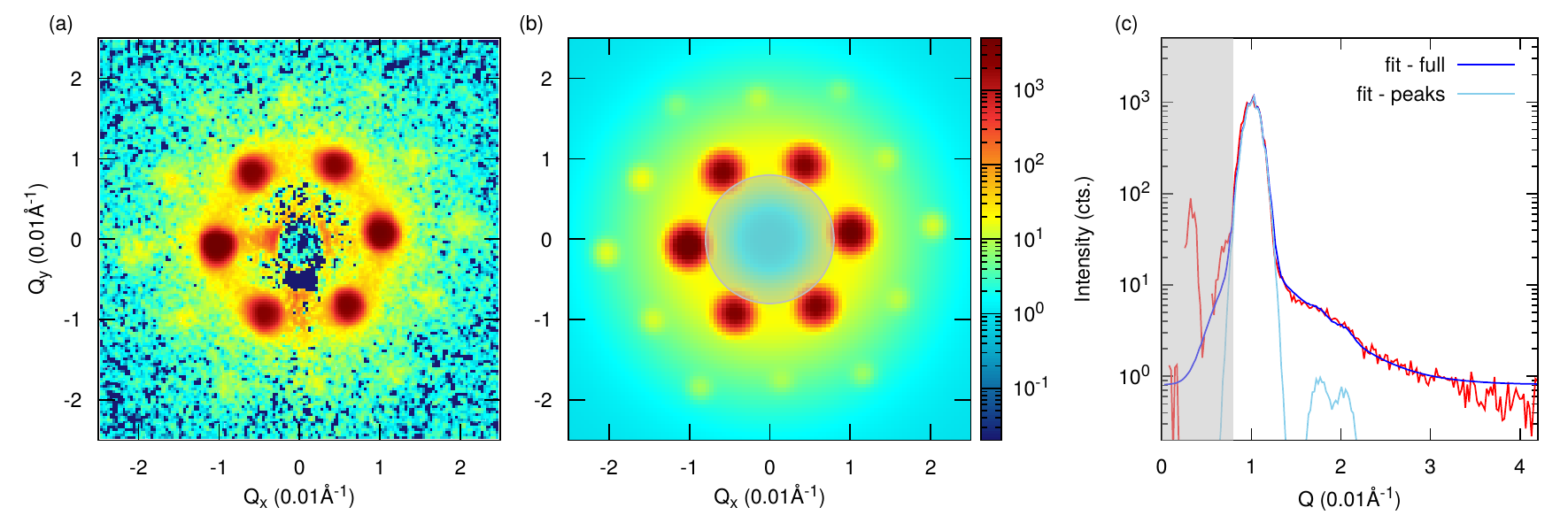}%
\caption{\label{fig:sup_map_94676}
  2D SANS intensity map obtained at $T=56.25$\,K and $\mu_0 H=17.4$\,mT using FC protocol.
}
\end{figure}

\begin{figure}
\includegraphics{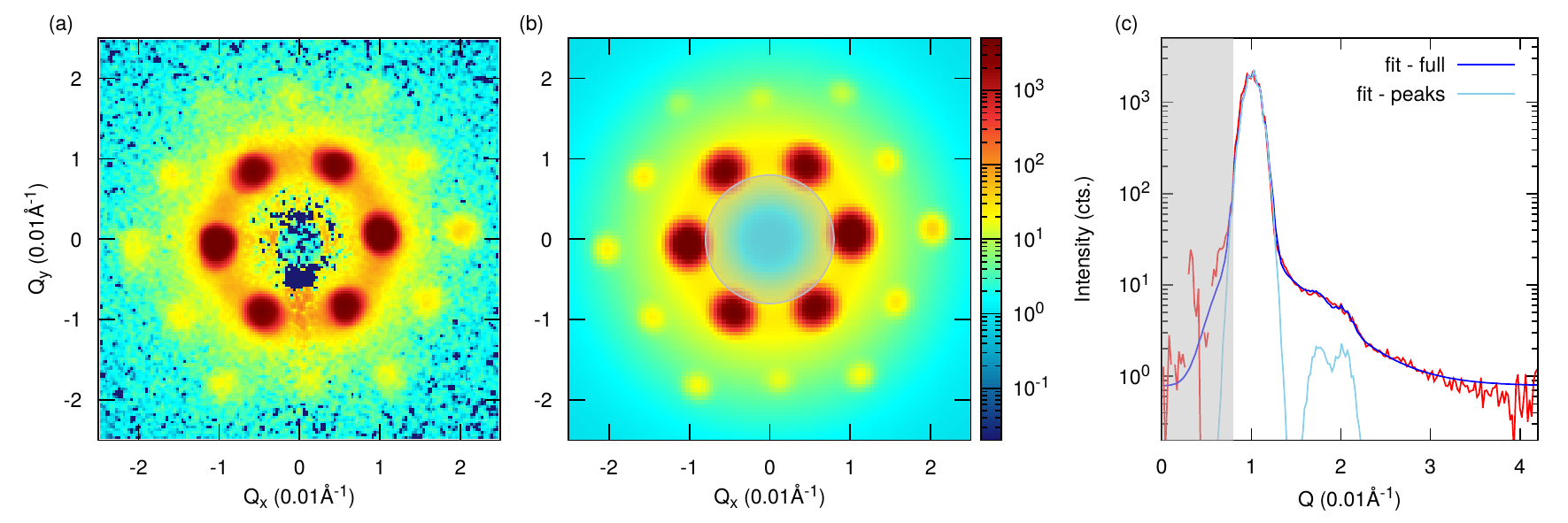}%
\caption{\label{fig:sup_map_95379}
  2D SANS intensity map obtained at $T=56.50$\,K and $\mu_0 H=17.4$\,mT using FC protocol.
}
\end{figure}

\begin{figure}
\includegraphics{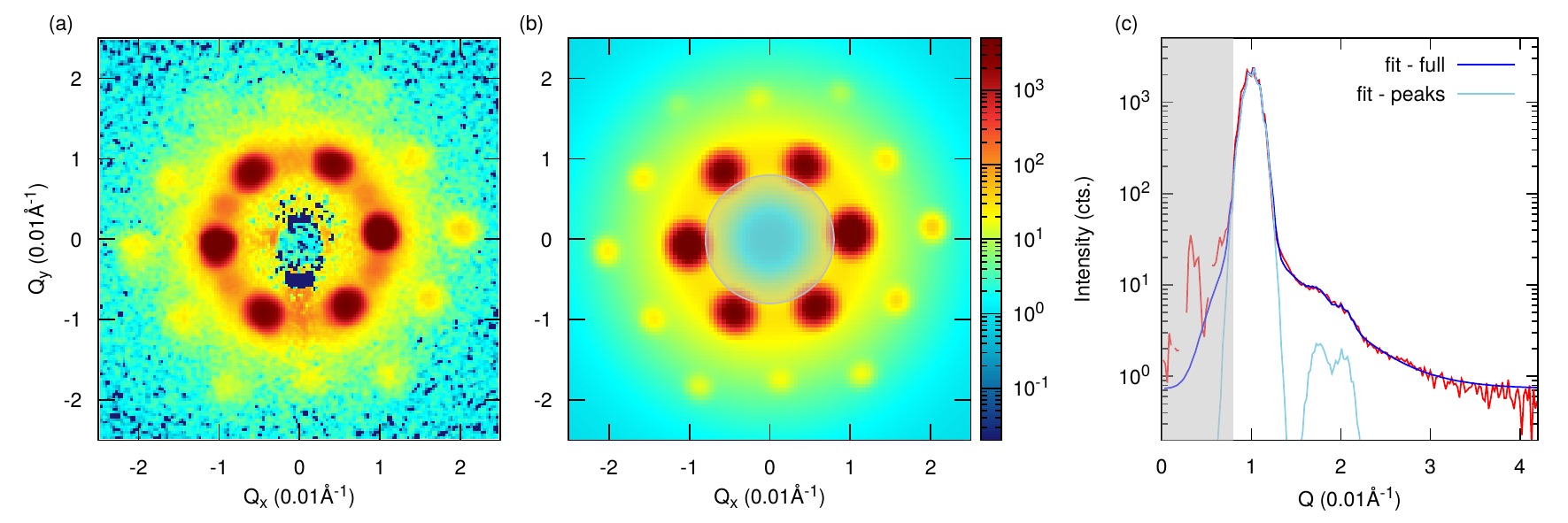}%
\caption{\label{fig:sup_map_95888}
  2D SANS intensity map obtained at $T=56.75$\,K and $\mu_0 H=17.4$\,mT using FC protocol.
}
\end{figure}

\begin{figure}
\includegraphics{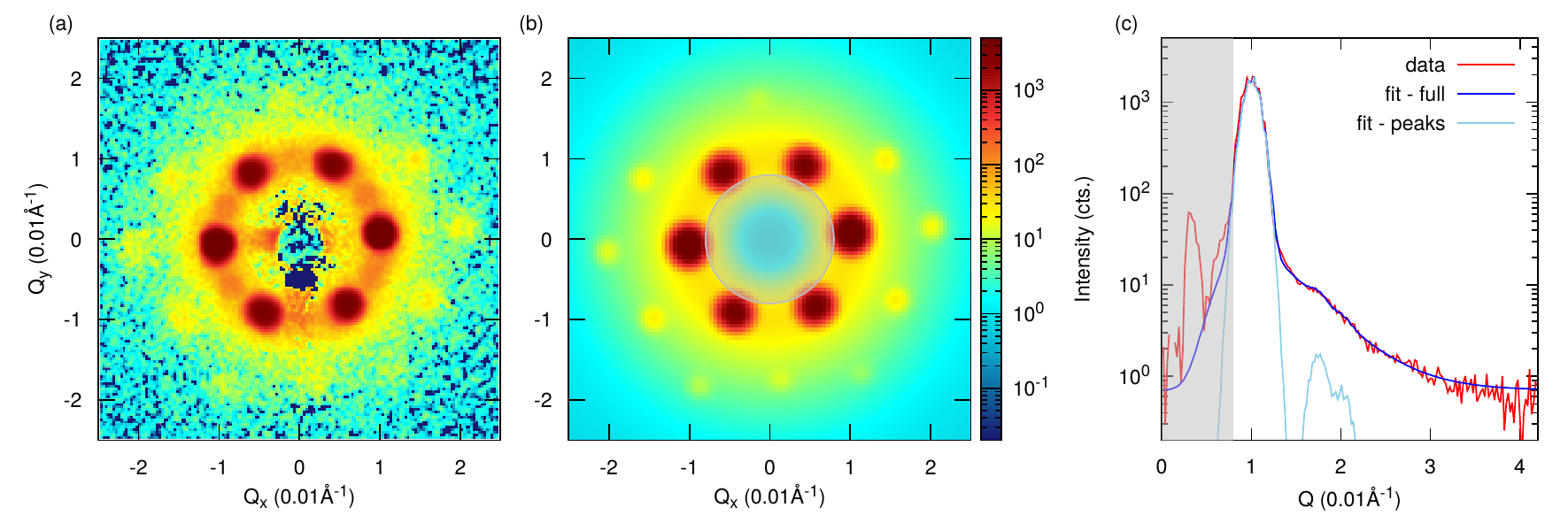}%
\caption{\label{fig:sup_map_94616}
  2D SANS intensity map obtained at $T=57.0$\,K and $\mu_0 H=17.4$\,mT using FC protocol.
}
\end{figure}

\begin{figure}
\includegraphics{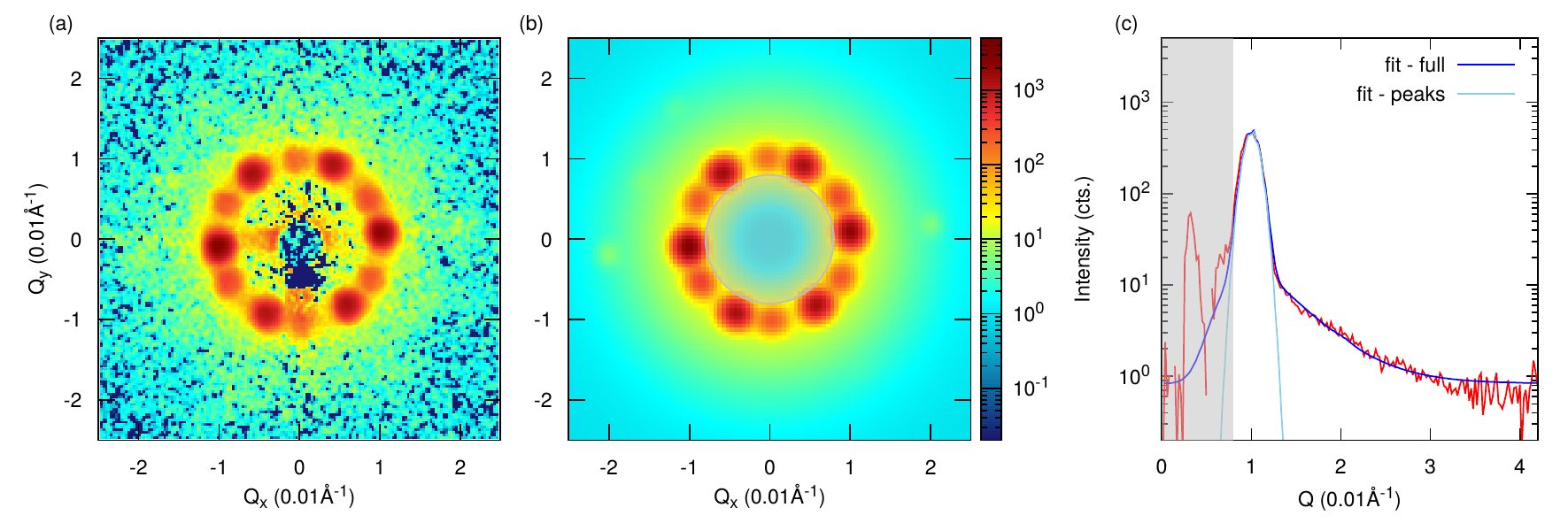}%
\caption{\label{fig:sup_map_94766}
  2D SANS intensity map obtained at $T=56.75$\,K and $\mu_0 H=13.4$\,mT using ZFC protocol.
}
\end{figure}

\begin{figure}
\includegraphics{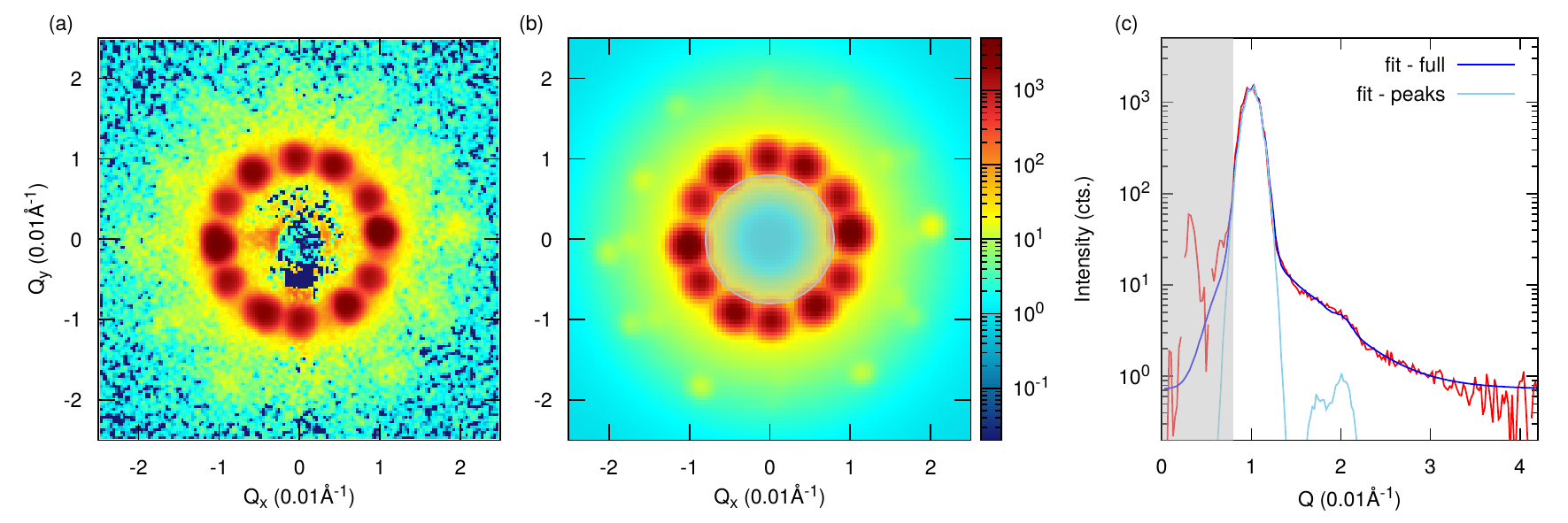}%
\caption{\label{fig:sup_map_94796}
  2D SANS intensity map obtained at $T=56.75$\,K and $\mu_0 H=15.4$\,mT using ZFC protocol.
}
\end{figure}

\begin{figure}[ht!]
\includegraphics{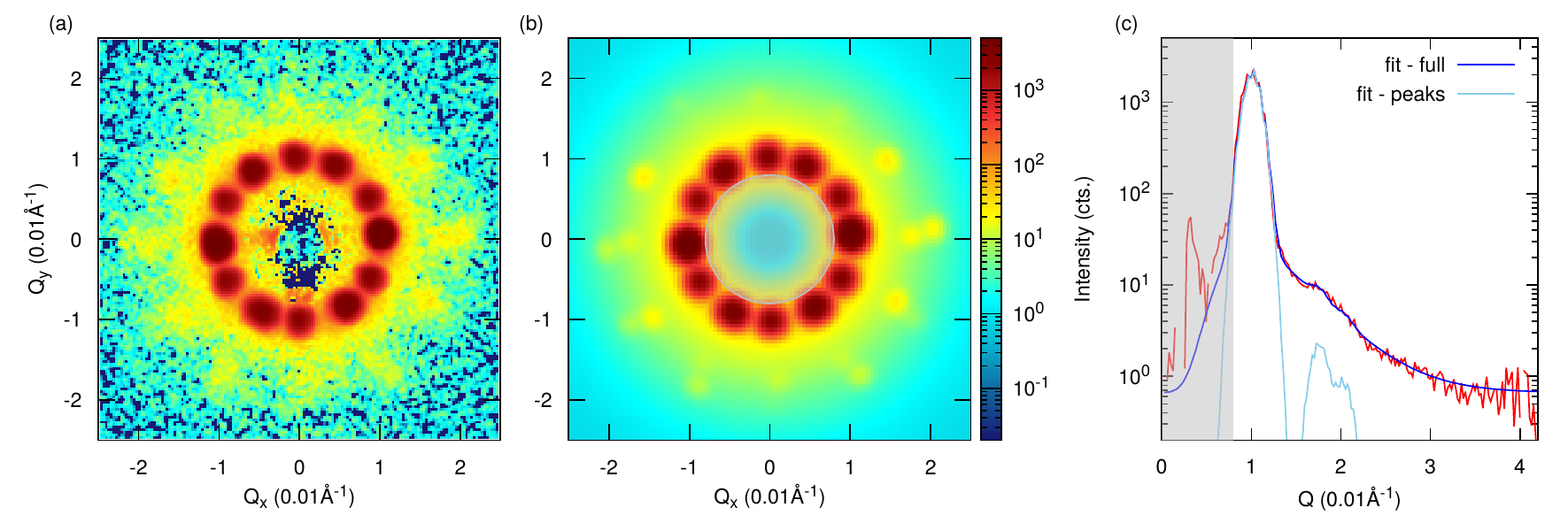}%
\caption{\label{fig:sup_map_94856}
  2D SANS intensity map obtained at $T=56.75$\,K and $\mu_0 H=19.4$\,mT using ZFC protocol.
}
\end{figure}

%

%

%

%

%

%